\documentclass[10pt]{article}
\usepackage{cite}
\usepackage{multicol}
\usepackage{graphicx}
\usepackage{amsmath}
\usepackage[a4paper]{geometry}
\usepackage{hyperref}
\usepackage{rotating}
\usepackage{xcolor}

\newcommand{\la}{\langle}
\newcommand{\ra}{\rangle}

\begin{document}

\begin{center}
{\Large \bf Thermodynamic Analysis of Transverse Momentum Spectra in Pb-Pb \vspace*{1mm}\\  Collisions at 2.76 TeV: Centrality Dependence of Temperature,\vspace*{2.4mm}\\
Freezeout Parameters and Non-Extensitivity} 
\vskip8mm
M. Waqas$^{1}${\footnote{Corresponding author: (M. Waqas)
waqas\_phy313@yahoo.com, 20220073@huat.edu.cn}},
Hassan Ali Khan$^{1}$, Wolfgang Bietenholz$^{2}$
{\footnote{wolbi@nucleares.unam.mx}},
Muhammad Ajaz$^{3}${\footnote{ajaz@awkum.edu.pk}},\\ 
Jihane Ben Slimane$^{4}$,
Haifa I.\ Alrebdi$^{5}$, A. Haj Ismail$^{6}$ \vspace*{2mm}
\\
{\small\it $^1$ Hubei Key Laboratory of Energy Storage and Power Battery, School of Optoelectronic Engineering \\ School of New Energy, Hubei University of Automotive Technology, Shiyan 442002, China\\
      $^2$ Instituto de Ciencias Nucleares, Universidad Nacional Aut\'onoma
      de M\'exico \\  Apartado Postal 70-543, CdMx 04510, Mexico\\
  $^3$ Department of Physics, Abdul Wali Khan University Mardan \\
  23200 Mardan, Pakistan\\
  $^4$ Department of Computer Sciences, Faculty of Computing and
  Information Technology \\ Northern Border University, Rafha, 91911,
  Saudi Arabia\\
  $^5$ Department of Physics, College of Science, Princess Nourah
  bint Abdulrahman University \\ P.O. Box 84428, Riyadh 11671,
  Saudi Arabia\\
  $^6$ College of Humanities and Sciences, Ajman University, Ajman P.O. Box 346, UAE\\}
\end{center}

{\bf Abstract:} 
We study properties of Pb-Pb collisions at 2.76 TeV in mid-rapidity,
$|y|<0.5$, based on data by the ALICE Collaboration.
In particular, we examine the transverse momentum ($p_T$) spectra of
positively charged (identified) hadrons, $\pi^+$, $K^+$ and $p$,
generated in various centrality intervals.
We perform individual fits using the 
thermodynamically consistent Tsallis distribution to extract the following
quantities: the non-extensitivity parameter, $q$, the effective
temperature, $T_{\rm eff}$, the kinetic freezeout volume, $V$, the mean
transverse flow velocity, $\beta_T$, the mean kinetic freezeout
temperature, \(\langle T_0\rangle\), the thermal temperature, $T_{\rm th}$,
and the parameter
$\zeta$, which characterizes the fluctuating number of generated particles.
From peripheral to central collision, and from lower to higher
charged particle multiplicity per pseudorapidity unit,
\(\langle dN_{\rm ch}/d\eta \rangle\), all these quantities are observed to
increase, with the exception of $q$, which has the opposite behavior.
The parameters $T_{\rm eff}$, $q$, and $V$ depend on the hadron mass in
a way that supports the scenarios of volume differential
freezeout and multiple kinetic freezeout. Furthermore,
  we extracted \(\langle T_{\rm eff}\rangle\) and \(\langle q\rangle\) for
  different collisions and energies at LHC and RHIC, and compare
  their dependencies on \(\langle dN_{\rm ch}/d\eta \rangle\) and
  \(\langle N_{\rm part} \rangle\).

{\bf Keywords:} Quark-Gluon Plasma, effective temperature, transverse
    flow velocity, kinetic freezeout temperature, kinetic freezeout volume,
charged particle multiplicity

\vskip1.0cm

\begin{multicols}{2}

{\section{Introduction}}
In heavy ion collisions, the Quark-Gluon Plasma (QGP) creation at sufficiently high temperatures
and densities are suggested by lattice Quantum Chromodynamics (QCD)
simulations \cite{QGPLQCD}.
The liberation of partonic degrees of freedom from the nucleons,
to form the thermalized QGP medium is thought to be the source of
this remarkable phenomenon in relativistic
nucleus-nucleus collisions. A substantial expansion flow takes
place during the parton evolution stage, before the system's temperature
drops below the threshold for parton-to-hadron conversion. This occurs
because deconfined partonic matter rapidly expands due to thermal
pressure against the surrounding vacuum \cite{Srivastava:2016ayf}.
The particle interactions end when the system is sufficiently diluted
due to additional re-scattering between the generated hadrons
\cite{Retiere:2003kf}. The final state
particle momentum distributions encode the medium's evolving
information, which is frequently described by relativistic fluid
hydrodynamics, see {\it e.g.}\ Refs.\ \cite{Gale:2013da,Heinz:2013th}.

In nuclear and particle physics, high-energy ion collisions are
a major field of research as they shed light on the characteristics
of strongly interacting matter in extreme environments \cite{Bjorken:1982qr}.
One experimental observable that provides valuable insight into the mechanisms
at play, at all event sizes that culminate in the final state of generated
particles is the Transverse Momentum ($p_T$) Distribution (TMD).

The TMD can be illustrated by a histogram constructed with the $p_T$
values of the emerging charged
particles per momentum space unit \cite{aaa}. Because of the TMD's
significance, theoretical research and empirical
fits --- that sufficiently characterize a portion or the
entirety of the spectrum --- are required. The temperature of the collision
system has often been linked to the inverse of the exponential decline in
previous attempts to parameterize this distribution,
which assumed the TMD to have an exponential form
\cite{Becattini:1997uf,Becattini:1995if}.
At low center-of-mass energies, this fitting function could explain
the experimental results fairly well. This methodology is appropriate
in cases where soft scattering mechanisms account for the majority
of the spectral contribution, resulting in a soft thermal distribution
akin to $p_T$
\cite{Feal:2020myr,Braun-Munzinger:1994ewq,Braun-Munzinger:2003htr}.
As the experiments proceeded towards higher energies, however, a
non-exponential tail was observed \cite{Feal:2020myr,Bialas:2015pla}.

In the early 1980s, Hagedorn suggested a QCD-based fitting function
\cite{Hagedorn:1983wk,Hagedorn:1967dia,Hagedorn:1964zz}. This function
was defined by a power-law in $p_T$, adjusted by a threshold that
had been determined by the momentum scale of elastic scattering.
It is noteworthy that this approach was able to replicate the
thermal response of the TMD and the power-law tail at high and low $p_T$.
The high-energy physics community later proposed a new fitting
function that extends the thermal distribution by incorporating
some non-extensitivity characteristics of the systems produced
in high-energy collisions \cite{Wilk:1999dr,Biro:2020kve}. Its
foundation lies in the Tsallis $q$-exponential function. However,
it has been demonstrated that these fitting functions are indeed
equivalent\cite{Saraswat:2017kpg}.

Free-fall scenarios involving chemical and kinetic freezeout
primarily characterize the evolution generated in relativistic
heavy-ion collisions. A definite yield of generated particles is
reached during the chemical freezeout stage, which is represented
by the cessation of inelastic collisions between hadrons and the
subsequent composition of new bound states. A shift in the particle
momenta occurs at the kinetic freezeout, as a result of the continuous
elastic interactions between the generated particles. The system remains
in chemical and kinetic equilibrium until its abrupt freezeout,
and it is believed that they occur simultaneously at the transition
between hadronic phase and QGP \cite{Broniowski:2001we}.
An alternative scenario that involves kinetic and chemical freezeouts
occurring at notably different times can also be considered. Elastic
collisions between hadrons would cause the resonances created at
the chemical freezeout stage to fade more quickly, and the system to
evolve. During this phase, the system maintains local thermal equilibrium
until it approaches kinetic freeze-out \cite{Retiere:2003kf}.

Systematic investigations have been conducted into the centrality
dependence of the kinetic freezeout properties (kinetic freezeout
temperature and radial flow)
\cite{Che:2020fbz,Waqas:2024qlo,Waqas:2019mjp,STAR:2017sal,Waqas:2021qkr,Wang:2023rpd,Sharma:2024nkt,ALICE:2013mez,CMS:2016zzh,ALICE:2013wgn}.
The collective flow arises from the pressure gradient during the
partonic system's expansion phase. This results in a
unique dependence of the form of the $p_T$ distribution on the particle
mass, which may be explained using a shared transverse
flow velocity and kinetic freezeout temperature \cite{Schnedermann:1993ws}.
It is therefore one important approach to assess the $p_T$ distribution of
hadrons with a global fit. However, the global fitting is not useful to
explore the open issue of the freezeout scenario of the particle,
for instance the single, double or multiple kinetic freezeout scenarios.

In this work, we choose the individual fitting to the $p_T$ distribution
to individually study the properties of three particle types, which
is useful in unraveling the correct freezeout scenario.
We investigate the freezeout parameters,
which are extracted from data by the ALICE Collaboration working at the
Large Hadron Collider (LHC) at CERN. We first extract the effective
temperature, $T_{\rm eff}$, then the kinetic transverse flow velocity and
freezeout temperature are obtained from $T_{\rm eff}$ by using an
alternative method. It should be noted that different methods for
extracting these parameters have different scales, therefore it is
not surprising if the explicit values of these quantities
are somewhat different.

A variety of models and distributions have been used for the
investigation of the freezeout parameters in a number of studies
\cite{Hagedorn:1983wk,Schnedermann:1993ws,STAR:2006nmo,UA1:1982fux,Cleymans:2011in,Pereira:2007hp,Conroy:2010wt}. In this work, we choose the Tsallis
distribution, because nucleus-nucleus collisions at high energies often
exhibit a non-extensive behavior. For our study, we use the form
of Tsallis distribution of Ref.\ \cite{Cleymans:2011in}, in
which the thermodynamic consistency is correctly implemented.

Section II provides an example of how the Tsallis distribution approach is put
into practice. Section III presents our results for the freezeout parameters
and further observables, which are summarized and discussed in Section IV.
 
{\section{The method and formalism}} 
Much research is done on the TMD of the departing hadrons in
high-energy collisions. The spectral form of the distributions
can be described by the standard exponential distributions in the
low-$p_T$ regime. The following formula is based on the assumption
that the chemical potential is negligible at high energies,
\begin{eqnarray}
\label{eq1}
f(p_T) \approx \exp \bigg[- \frac{m_T(p_T)}{T}\bigg] .
\end{eqnarray}
Here $T$ is the appropriate temperature, and $m_T$ is the ``transverse
mass'', defined as $m_T = \sqrt{p_T^2+m^2}$, where $p_T$
is the transverse momentum and $m$ is the particle's (rest) mass.

The particles in the high-$p_T$ region are usually described
by power-law distributions, which are frequently used in high-energy
physics \cite{Hagedorn:1983wk,ALICE:2013txf,CMS:2010tjh,ATLAS:2010jvh,ALICE:2010syw,PHENIX:2011rvu}
\begin{align}
  f(p_T) = \frac{1}{N} \frac{\mathrm{d}N}{\mathrm{d}p_T} =
  A \, p_T \left( 1 + \frac{p_T}{p_0} \right)^{-n} ,
\end{align}
where $N$ represents the particle number, $A$ is the normalization
constant, while $p_0$ and $n$ are treated as free parameters.

The Tsallis distribution is widely used in the community with different
versions which covers the high-$p_T$ region. 
In the present study, the non-extensive statistics of Tsallis' distribution
is identical to that of Ref.\ \cite{Cleymans:2011in},

\begin{eqnarray}
\label{eq3}
\frac{d^2N}{dp_T\ dy} &=& g \, V \, \frac{p_T\ m_T \cosh y}{(2\pi)^2} \nonumber \\
&& \times\bigg[1+(q-1)\frac{m_T \cosh y - \mu}{T}\bigg]^{q/(1-q)} .
\ \quad
\end{eqnarray}
The parameter $g$ is the degeneracy factor, $V$ is the kinetic
freezeout volume, $q$ is the non-extensitivity parameter
(or entropic index), $y$ the rapidity and $\mu$ the chemical potential.

Some studies found a link between the parameters $q$ and $T$
\cite{Wilk:1999dr,Biro:2012fiy,Wilk:2012zn}. The so-called Tsallis
thermometer in the $T$-$q$ diagram can be used to measure this
correlation.
It was demonstrated that the total charged hadron multiplicity,
which follows the Negative Binomial Distribution (NBD), can be
explained by the Tsallis distributed transverse momentum. Experimental
data confirm this as well, yielding an NBD value of $k \sim {\rm O}(10)$
where $k$ is related to the non-extensitivity as $k = (2-q)/(q-1)$
\cite{ATLAS:2010jvh,PHENIX:2008psu}. This observation leads to an
explanation that takes into consideration the variations $\Delta N$
in the number of created particles, $N$, in a $1+1$-dimensional
relativistic gas 
\cite{Biro:2014fna},
\begin{equation}
\label{eq4}
T = \frac{E}{\langle N \rangle }, \quad
q = 1 - \frac{1}{\langle N \rangle} + \frac{\triangle N^2} {\langle N \rangle^2} .
\end{equation}
Numerous recent studies have demonstrated the crucial and intriguing
role of event-by-event multiplicity measurements and, consequently,
their variations in high-energy physics.
The evolution of long-range correlations in high multiplicity
$p$-$p$ and $p$-Pb events, or the continuous enhancement of (multi-)strange
hadrons in relation to multiplicity
\cite{Grosse-Oetringhaus:2019axb,ALICE:2016fzo,ATLAS:2012cix,Mishra:2018pio,Nath:2019mkf,Zaccolo:2015udc}, are two examples.

Our work focuses on the values of the parameters $T$ and $q$ in different
centrality classes. Assuming the relative scale of multiplicity fluctuations
to be constant, as in Ref.\ \cite{Biro:2020kve}, the relation
between $T$ and $q$ in the thermodynamic picture can be expressed
in terms of the parameter $\zeta$,
\begin{eqnarray}
\label{eq5}
\zeta^2 = \frac{\triangle N^2}{\langle N \rangle^2} .
\end{eqnarray}
By referring to Eq.\ (\ref{eq4}), we obtain
\begin{eqnarray}
\label{eq6}
T = E \left[ \zeta^2-(q-1) \right] .
\end{eqnarray}

\begin{figure*}
\centering
  \includegraphics[width=0.45\textwidth]{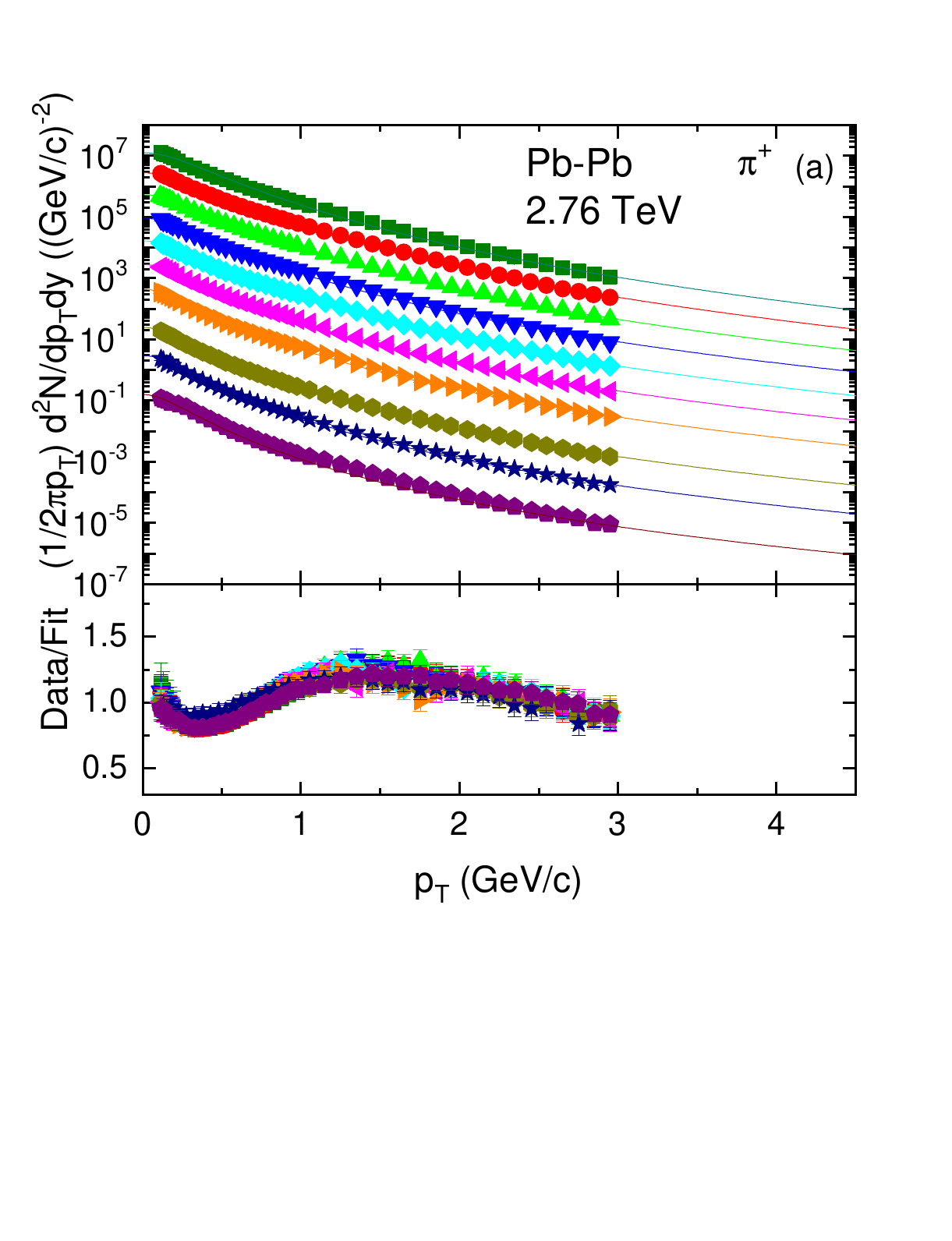}
 \includegraphics[width=0.45\textwidth]{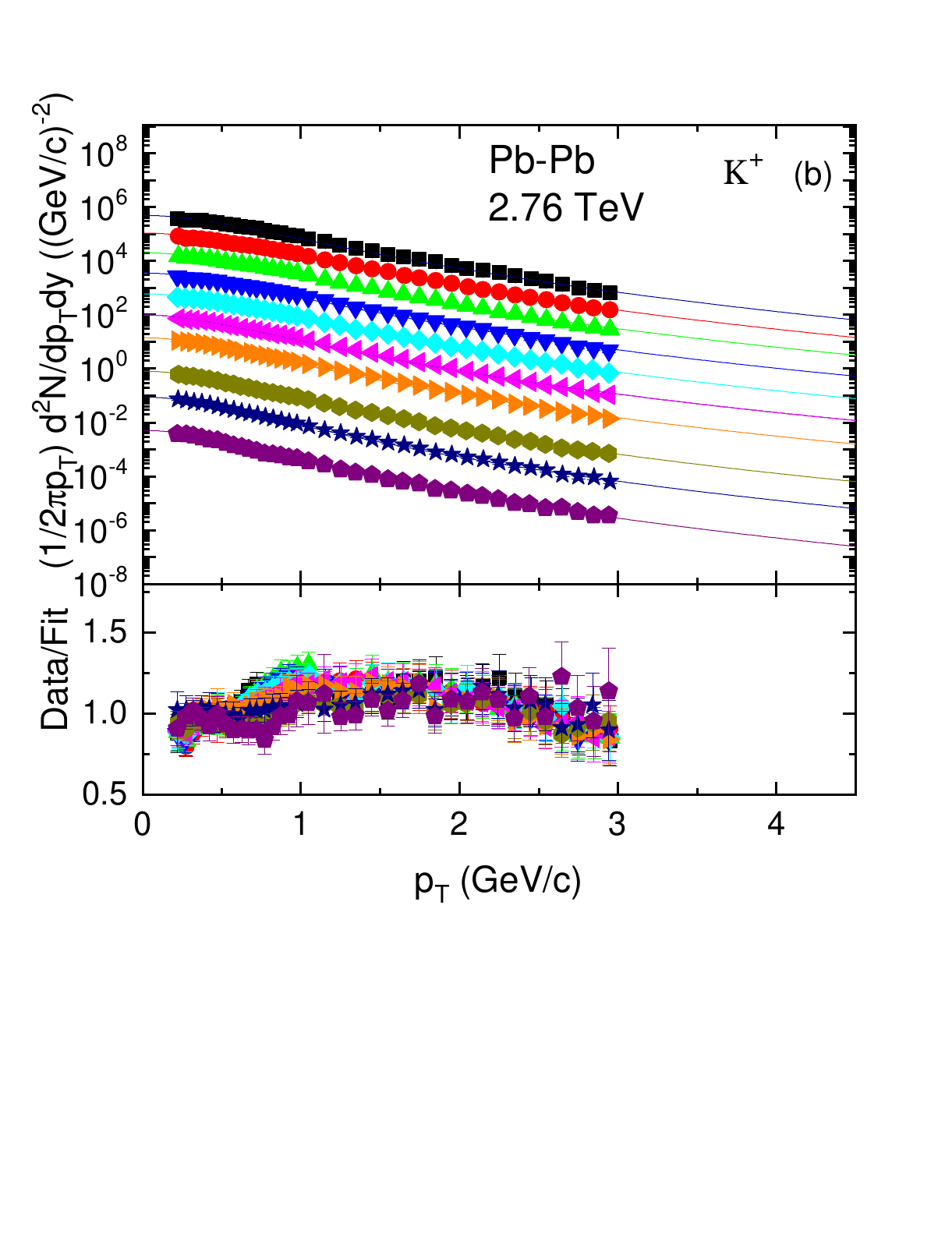}
 \includegraphics[width=0.45\textwidth]{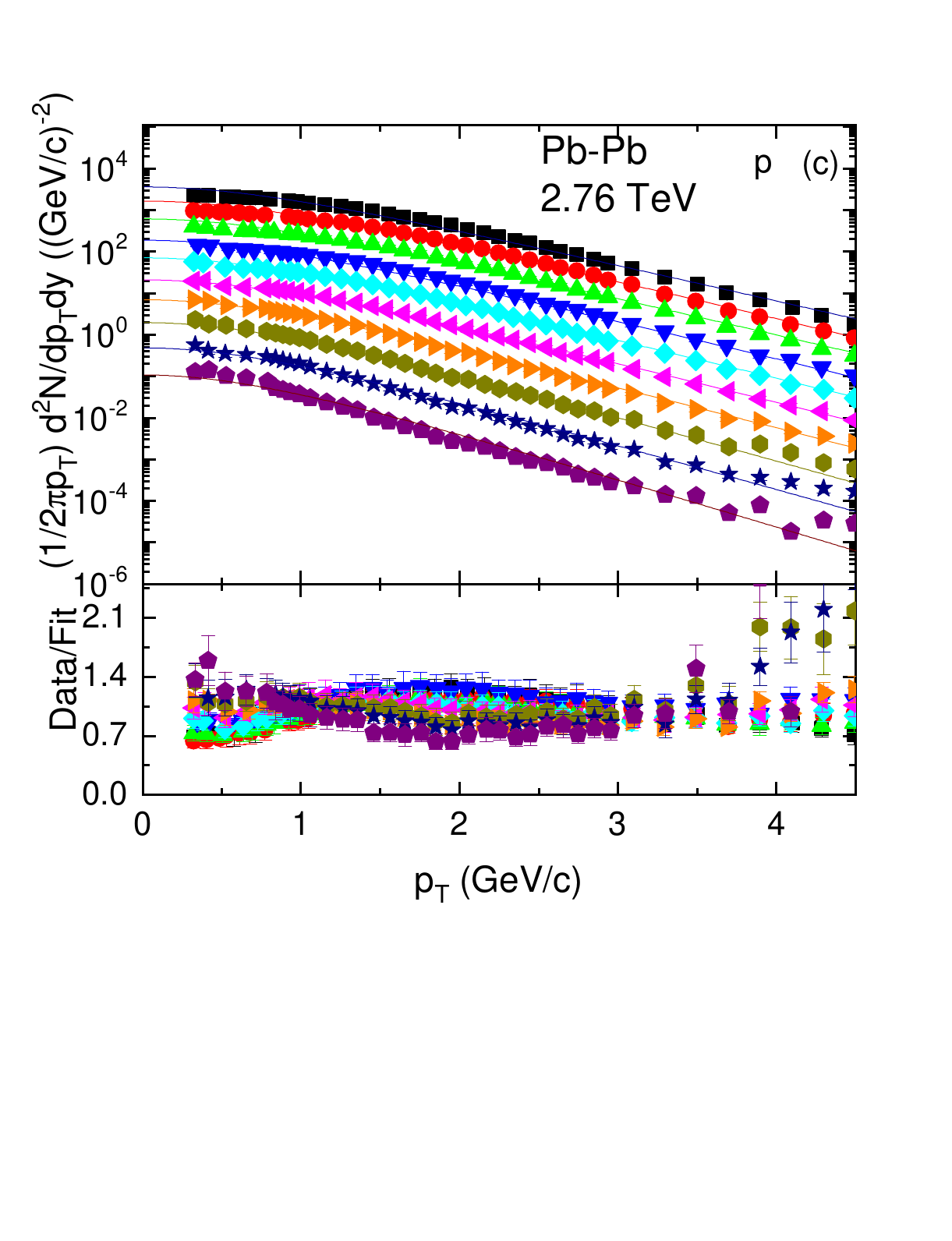}
\caption{The $p_T$ spectra of $\pi^+$, $K^+$, and $p$, generated
  in Pb-Pb collisions at $\sqrt{s_{NN}}=2.76~{\rm TeV}$ and
  mid-rapidity, $|y|<0.5$, with various centralities. The experimental
  data recorded by the ALICE Collaboration is
  represented by consistent symbols in each panel, which display distinct
  particle $p_T$ spectra dispersed in different centrality classes
  \cite{ALICE:2013mez}, cf.\ Table \ref{tab1}\,. The curves represent our
  fits using the Tsallis distribution. The corresponding
  data/fit ratio for each plot is shown in the lower panels.}
\label{fig1}
\end{figure*}

Therefore, identifying the thermodynamic regime in the data of
various centrality intervals of Pb-Pb collisions at 2.76 TeV is
the main goal of this endeavor, which we may be able to do by
extracting $T$ and $q$.

Assuming collective flow and thermalization as the starting point,
the phenomenological model can account for almost all observed
hadronic spectra \cite{Schnedermann:1993ws}. In order to interpret
the mass dependence of the effective temperature $T$
\cite{Csorgo:1994fg,Helgesson:1997zz,Waqas:2020ioh,Moore:2004tg}
as the existence of a radial flow, a Gaussian parameterization
has been applied. The transverse particle momentum grows in direct
proportion to its mass due to the radial flow velocity, which
is created by powerful nucleon-nucleon collisions between two
colliding nuclei and evolves in hadronic re-scatterings as well
as the QGP phase \cite{Schnedermann:1993ws,Csorgo:2001xm,PHENIX:2011rvu}.
Numerous models try to study the radial flow \cite{Wei:2016ihj}.
For the purposes of this investigation, we can employ another
radial flow image \cite{Schnedermann:1993ws,PHENIX:2011rvu}
\begin{eqnarray}
\label{eq1}
T = T_0 + m \, \langle u_t \rangle^2 .
\end{eqnarray}
The variable $u_t$ represents the strength of the (mean radial)
transverse flow, and $T_0$ is the hadron kinetic freezeout
temperature. The relation between $u_t$ and the mean transverse
velocity $\langle \beta_T \rangle$ is given as
\begin{eqnarray}
\label{eq1}
\langle \beta_T \rangle = \frac{\langle u_t \rangle}{\sqrt{1 + \langle u_t \rangle^2}} .
\end{eqnarray}

{\section{Results and discussion}}
Using the thermodynamically consistent Tsallis distribution,
the observed $p_T$ distributions of positively charged, identified
hadrons, namely $\pi^+$, $K^+$ and $p$, in Pb-Pb collisions from
the ALICE collaboration, are shown and fitted in Figure \ref{fig1}.
Overall, there is a favorable match between the fit results and the
data points. Moreover, the identified particle's $p_T$ spectra in various
centrality intervals in Pb-Pb collisions at 2.76~TeV are fitted
to the same Tsallis distribution.

The panels in Figure \ref{fig1} represent the $p_T$ spectra of each
particle distributed in different centralities: panel (a), (b) and
(c) display the $p_T$ spectra of $\pi^+$, $K^+$, and $p$,
respectively. The data by the
ALICE Collaboration \cite{ALICE:2013mez} is represented by the
symbols in Figure \ref{fig1}, while the fits are displayed
by the corresponding curves. Various layers of $p_T$ distributions with
different symbols and colors are different centrality bins. The
data/fit ratio, which is depicted in the lower section of each panel,
is used to show the deviation of the fit from the experimental data.
The parameters extracted
from the model, as well as the values of $\chi^2$ and the
number of degrees of freedom (dof) are tabulated in Table \ref{tab1}.
$N_0$ in \ref{tab1} is the normalization constant which is used to
compare the fit result with the experimental data. Mathematically,
$N_0$ can be expressed as
\begin{align}
N_0 = \frac{1}{\int_{\rm min}^{\rm max} f(p_T) \, dp_T}
\end{align}
where the $p_T$ spectra are described by the model function $f_{p_T}$.
Another kind of normalization constant $C$, which is used
to normalize Eq.\ (\ref{eq3}), is expressed as $C =gV/(2\pi )^2$.

It is important that the fitting method we
  employed offers complementary physical insights that are equally
  valuable, even though the combined fit in Ref.\ \cite {ALICE:2013mez}
  undoubtedly provides the most direct evidence for a collective
  behavior by enforcing a common freezeout temperature and transverse
  flow velocity for all the particle species. The capacity of our
  individual fit analysis to test, rather than presume, a single
  freezeout scenario is what gives it its essential relevance.
  The freezeout situation is actually still up for debate
  \cite {Tang:2008ud,Chatterjee:2014lfa,Chatterjee:2015fua,Thakur:2016boy,Waqas:2020ygr,Waqas:2021rmb,imran2025}.
  By using the Tsallis distribution to evaluate each particle
  species independently, we can assess whether or not all of the
  particles actually conform to a single freezeout picture, or if
  there are small but physically significant variances that could
  indicate more intricate processes in the dynamical system. The fact
  that distinct particle species may undergo different degrees of
  thermalization or interact differently with the hadronic medium
  as a result of variations in mass, cross-section, or quark content
  makes this method especially pertinent. 

  A single freezeout scenario has been investigated in Ref.\
  \cite{ALICE:2013mez,STAR:2017sal,ALICE:2013wgn,ALICE:2019hno},
  which evaluated the identified particles by the blast wave model
  and obtained the freezeout parameters.
  In these studies, the system is assumed to be in local thermal equilibrium.
  We consider it favorable to employ a distribution like Tsallis
  that accounts for the non-equilibrium effects because we know
  that the system formed by heavy ion collisions at higher energy
  may well be out of local equilibrium. In the current work, the
  individual fit helps us to assess whether certain particles exhibit
  features that differ from the collective behavior. This could
  suggest phenomena like distinct freezeout times, non-thermal
  production mechanisms, or interactions specific to a species.
  Additionally, the validity of the collective freezeout picture
  is actually strengthened if the separately retrieved parameters
  are well aligned across various particle species; if they are
  not, this would reflect interesting physics beyond the most
  basic thermal scenario. This allows for a more nuanced knowledge
  of the freezeout process by using the individual fit strategy
  as a useful cross-check and modification of the simultaneous fit.

\begin{table*}
  \label{tab1}
{\scriptsize
\caption{Values of $T$, $q$, $V$, $N_0$ [which is the normalization constant to that compares the fit curve with experimental data], and
 $\chi^2$ per degree of freedom (dof) corresponding to the curves
  in Figure \ref{fig1}. All data refer to Pb-Pb collisions at
  2.76~TeV.} \vspace{-.4cm}
\begin{center} \label{tab1}
\begin{tabular}{ccccccccccc}\\ \hline\hline
 Centrality  & Particle   & $T~({\rm GeV})$
  & $q$ & $V~({\rm fm}^3)$   &    $N_0$        & $\chi^2$/dof \\
\hline
  $0-5\%$       & $\pi^+$ &$0.110\pm0.004$  & $1.132\pm0.005$  & $2380\pm66$    & $12500\pm171$      & 98/38\\
  $5-10\%$      & ---              &$0.105\pm0.003$  & $1.138\pm0.006$  & $2259\pm57$   & $10500\pm100$      & 101/38\\
  $10-20\%$    & ---              &$0.100\pm0.004$  & $1.144\pm0.004$  & $2144\pm81$   & $7592\pm144$       & 81/38\\
  $20-30\%$    & ---              &$0.096\pm0.004$  & $1.151\pm0.007$  & $2047\pm62$   & $5033\pm212$       & 74.5/38\\
  $30-40\%$    & ---              &$0.091\pm0.005$  & $1.155\pm0.008$  & $1965\pm48$   & $3351\pm153$       & 55.6/38\\
  $40-50\%$    & ---              &$0.086\pm0.004$  & $1.159\pm0.007$  & $1873\pm55$   & $2123\pm97$        & 82.4/38\\
  $50-60\%$    & ---              &$0.082\pm0.004$  & $1.162\pm0.008$  & $1759\pm50$   & $1235\pm111$       & 48.3/38\\
  $60-70\%$    & ---              &$0.078\pm0.003$  & $1.165\pm0.007$  & $1662\pm46$   & $657\pm52$         & 78.8/38\\
  $70-80\%$    & ---              &$0.075\pm0.003$  & $1.168\pm0.007$  & $1555\pm72$   & $300\pm29$         & 63.4/38\\
  $80-90\%$    & ---              &$0.071\pm0.003$  & $1.171\pm0.007$  & $1433\pm60$   & $110\pm18$         & 41.6/38\\
\hline
  $0-5\%$       & $K^+$    &$0.170\pm0.005$  & $1.120\pm0.006$  & $1729\pm85$   & $1550\pm110$      & 60/33\\
  $5-10\%$      & ---              &$0.164\pm0.005$  & $1.124\pm0.004$  & $1662\pm82$   & $1300\pm121$      & 92/33\\
  $10-20\%$    & ---              &$0.159\pm0.004$  & $1.131\pm0.007$  & $1650\pm74$   & $1000\pm97$       & 92/33\\
  $20-30\%$    & ---              &$0.154\pm0.005$  & $1.133\pm0.007$  & $1560\pm59$   & $680\pm143$       & 84/33\\
  $30-40\%$    & ---              &$0.147\pm0.003$  & $1.134\pm0.005$  & $1412\pm62$   & $440\pm49$        & 77/33\\
  $40-50\%$    & ---              &$0.142\pm0.003$  & $1.136\pm0.009$  & $1341\pm53$   & $280\pm24$        & 62/33\\
  $50-60\%$    & ---              &$0.138\pm0.004$  & $1.137\pm0.005$  & $1255\pm48$   & $155\pm31$        & 42.6/33\\
  $60-70\%$    & ---              &$0.134\pm0.004$  & $1.138\pm0.006$  & $1134\pm41$   & $83\pm28$         & 71/33\\
  $70-80\%$    & ---              &$0.128\pm0.004$  & $1.138\pm0.008$  & $1069\pm51$   & $35\pm8.1$        & 14.5/33\\
  $80-90\%$    & ---              &$0.122\pm0.004$  & $1.130\pm0.003$  & $1000\pm42$   & $14\pm2.7$        & 22/33\\
\hline
  $0-5\%$       & $p$ &$0.370\pm0.005$  & $1.036\pm0.007$  & $1212\pm52$   & $800\pm59$        & 60/33\\
  $5-10\%$      & ---              &$0.363\pm0.006$  & $1.038\pm0.006$  & $1156\pm69$   & $464\pm77$        & 92/33\\
   $10-20\%$    & ---              &$0.355\pm0.005$  & $1.043\pm0.008$  & $1098\pm71$   & $320\pm25.5$      & 92/33\\
   $20-30\%$    & ---              &$0.348\pm0.006$  & $1.045\pm0.008$  & $1025\pm66$   & $210\pm32.1$      & 84/33\\
   $30-40\%$    & ---              &$0.341\pm0.005$  & $1.048\pm0.007$  & $968\pm44$    & $144\pm32$        & 77/33\\
   $40-50\%$    & ---              &$0.335\pm0.005$  & $1.049\pm0.004$  & $911\pm57$    & $95\pm15.6$       & 62/33\\
   $50-60\%$    & ---              &$0.330\pm0.005$  & $1.050\pm0.006$  & $833\pm31$    & $53\pm7.4$        & 42.6/33\\
   $60-70\%$    & ---              &$0.324\pm0.006$  & $1.050\pm0.007$  & $755\pm28$    & $29\pm5.2$        & 71/33\\
   $70-80\%$    & ---              &$0.316\pm0.004$  & $1.041\pm0.005$  & $672\pm32$    & $15\pm3.8$        & 14.5/33\\
   $80-90\%$    & ---              &$0.310\pm0.005$  & $1.021\pm0.008$  & $600\pm31$    & $6\pm0.4$         & 22/33\\
\hline
\end{tabular}%
\end{center}}
\end{table*}

{\bf 3.1: Kinetic freezeout parameters}

In this subsection, we present the freezeout parameters, which we
obtained with the Tsallis fit. It should be noted that the
Tsallis distribution provides an excellent description over the
entire measured $p_T$ range for all three particle species. A combined
(global) fit to these species could offer important
information about the conditions under which the freezeout occurs.
On the other hand, various particle types can separate
from the system at different instances, resulting in distinct values
of $T_0$ and $\beta_T$ in the hadronic medium. That can be
investigated by individual fits to $p_T$ spectra.

\begin{figure*}
\centering
 \includegraphics[width=0.45\textwidth]{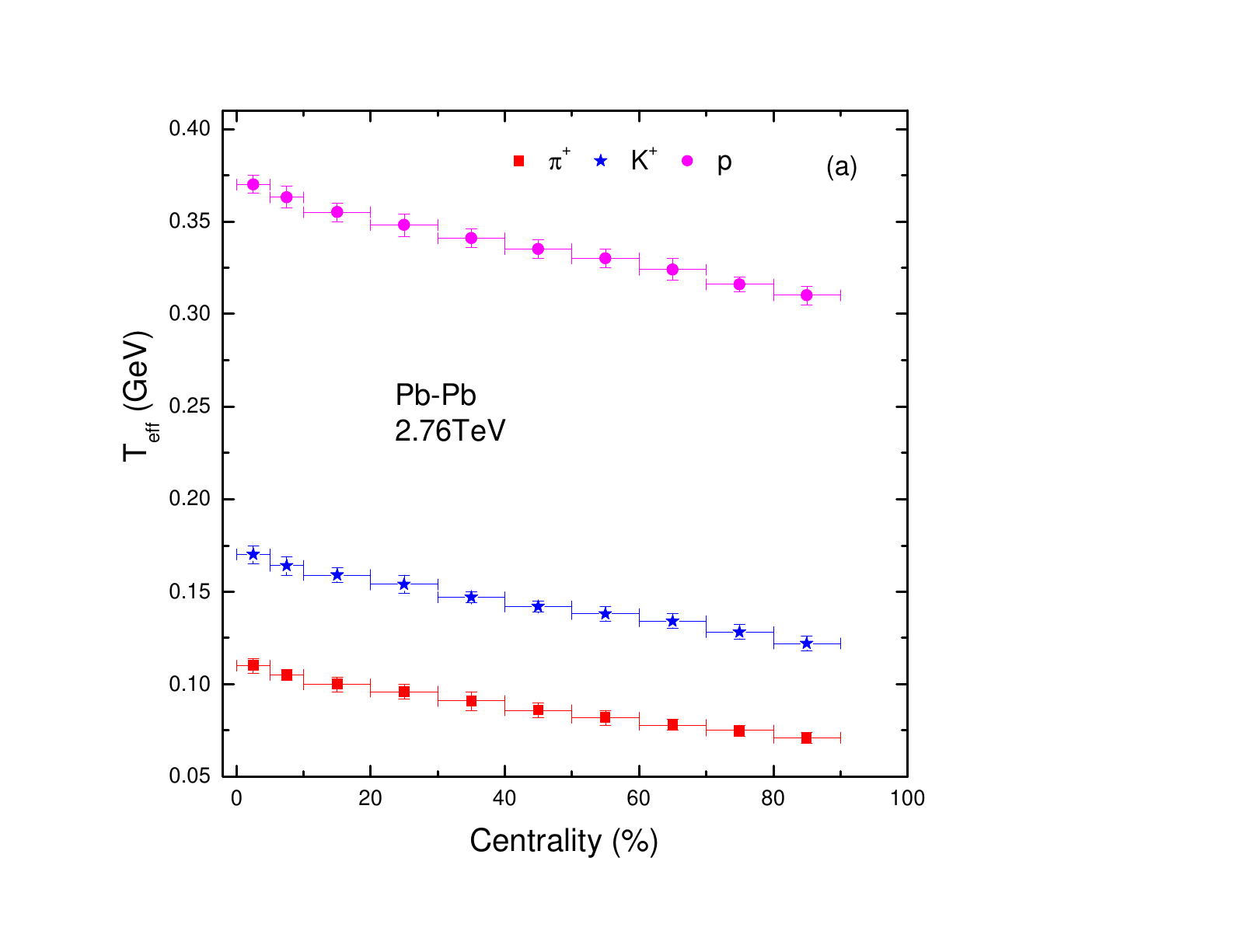}
 \includegraphics[width=0.45\textwidth]{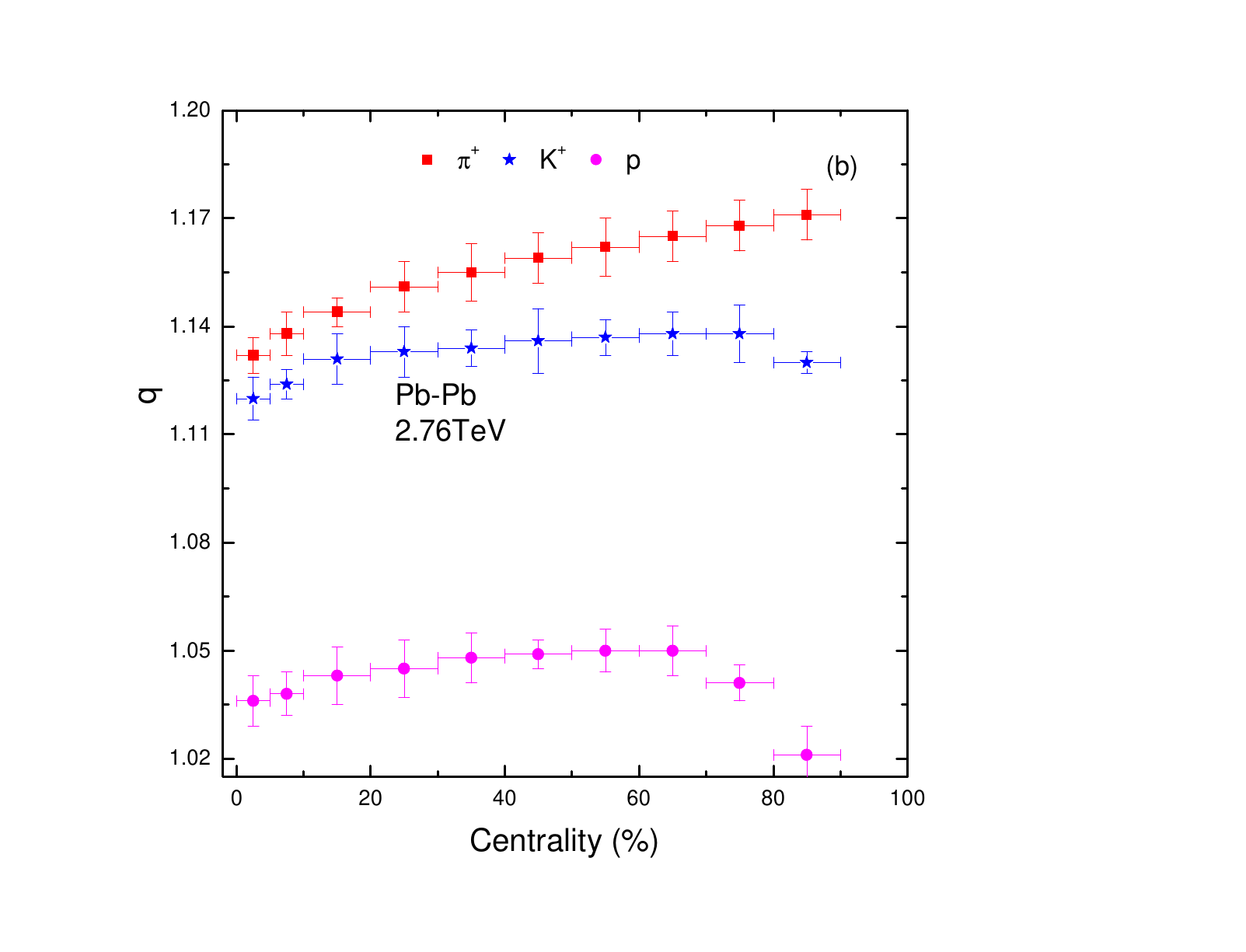}
 \includegraphics[width=0.45\textwidth]{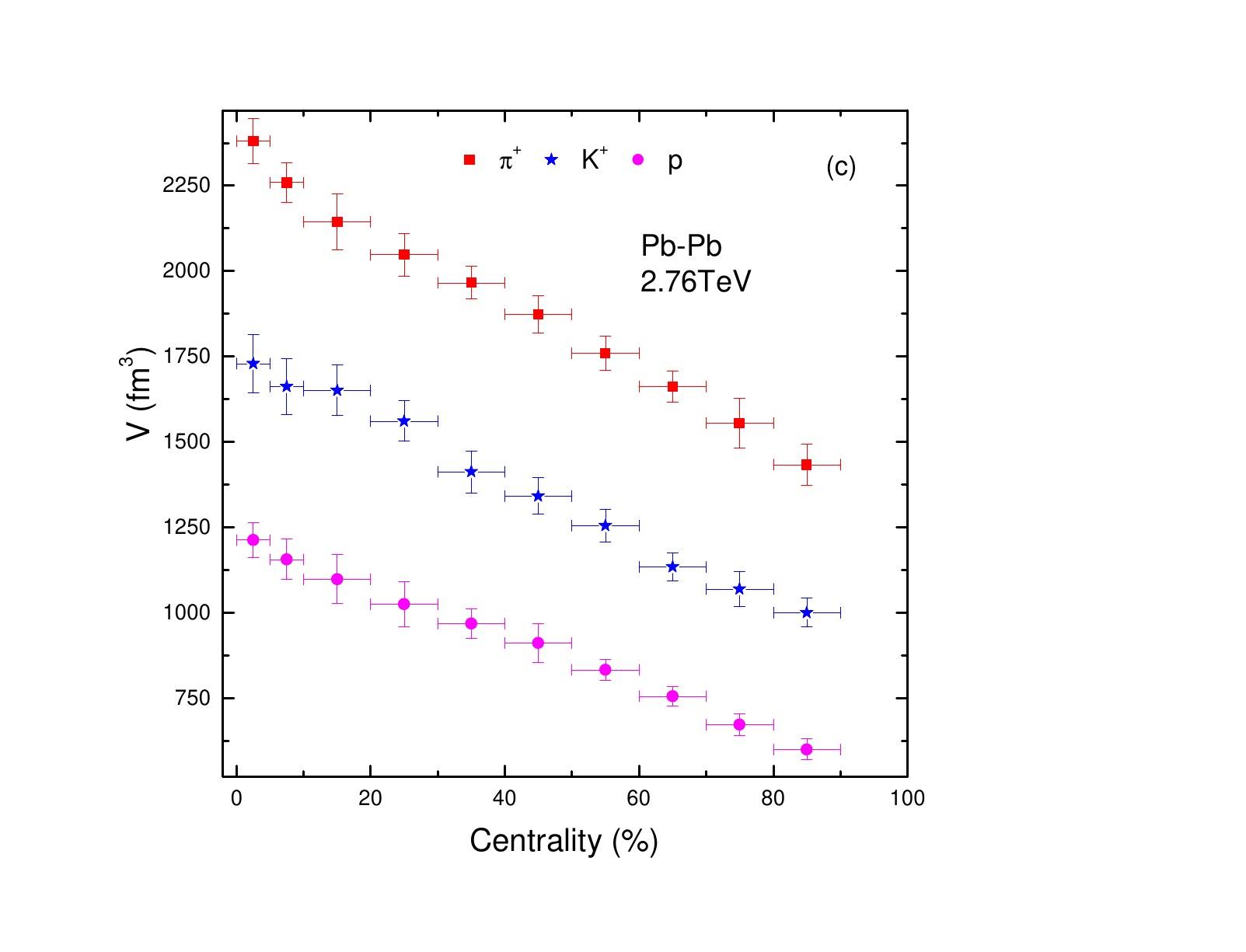}
\caption {Effects of centrality and the particle mass on (a) the
  effective temperature $T_{\rm eff}$,
  (b) the  non-extensitivity parameter $q$, and (c)
  the kinetic freezeout volume $V$.}
\label{fig2}
\end{figure*}

The centrality dependence of the extracted parameters is depicted
in Figure \ref{fig2}. Different panels display the result for different
parameters: panel (a), (b) and (c) show the behavior of the effective
temperature $T_{\rm eff}$, of the non-extensitivity parameter $q$,
and of the kinetic freezeout volume $V$, respectively, for different
centralities. Different symbols denote distinct particles in each
panel. In every panel, we recognize the dependence of the
corresponding parameter on the particle mass, while the horizontal
sequence of symbols illustrates the dependence on centrality.

In panel (a) of Figure \ref{fig2}, we see that $T_{\rm eff}$ is
reduced as we proceed from central to less central collisions,
with higher $T_{\rm eff}$ values for heavier particles.
Large numbers of participants in
the reaction deposit a huge amount of energy in central collisions,
increasing the system's degree of excitation and, consequently,
its $T_{\rm eff}$. This effect is most prominent for the proton,
the heaviest particle in this set, followed by $K^+$ and $\pi^+$.
In our previous work \cite{Waqas:2022nae}, we analyzed
$p_T$ of strange particles and obtained the same trend of the
temperature from central to peripheral collisions.

Additionally,
in the current work, $T_{\rm eff}$ depends on the masses, which amount
to $m_{\pi^{+}} \simeq 139.6 ~{\rm MeV}$,
$m_{K^{+}} \simeq 493.7 ~{\rm MeV}$, $m_{p} \simeq 938.3 ~{\rm MeV}$. Heavy particles are more likely to freezeout quickly, because of
their frequent interaction.
 Our results for the mass dependence of $T_{\rm eff}$ reveal
the multiple kinetic freezeout
situation \cite{Waqas:2018xrz,Waqas:2020ygr}. The mass
dependence of $T_{\rm eff}$ is most prominent for the protons.

Likewise, the dependence of non-extensitivity parameter $q$ on
centrality and $m$ is shown in panel (b). With respect to centrality,
the behavior of $q$ is opposite to $T_{\rm eff}$: it decreases
with the particle mass and increases from central to peripheral
collisions.
The value $q=1$ corresponds to the conventional Boltzmann-Gibbs
statistics. The system tends to be far from equilibrium
when $q$ differs significantly from 1.
We observe that $q > 1$ rises from central to peripheral
collisions, indicating that the system tends to get
out of equilibrium. For heavier particles, $q$ is
smaller, indicating that the heavy particles attain the
state of equilibrium faster. Indeed, the proton
$p$ is seen to have its $q$-parameter close to 1.
The variation of $q$ is most
pronounced in case of $\pi^+$, where it continuously grows
as we proceed from central to peripheral collisions. In case
of $K^+$, however, it remains almost
constant in peripheral collisions. For the proton, it first
increases towards peripheral collisions, and then --- in the
least central collisions --- it decreases again. This peculiar
behavior might be related to the process of baryon stopping
and to baryon number conservation.
When protons remain mostly as leftovers of the initial
colliding nuclei, projectile fragmentation may become the dominant
mechanism of $p$ creation in very peripheral collisions. This could
lead to a less fragmented or more thermally-like distribution of
protons, which would lower $q$.

Panel (c) displays the result for
the kinetic freezeout volume $V$. $V$ is seen to depend on both
the centrality and particle mass. It decreases from head-on to
peripheral collisions, because many particles participate in
head-on collisions. This means that many binary collisions result
from parton re-scattering, which swiftly returns the system to
equilibrium. In peripheral collisions, the number of participants
decreases as the system steadily approaches equilibrium.
Apart from these findings, $V$ is largest for $\pi^+$, followed
by $K^+$ and $p$. This enables a volume differential freezeout
and shows that each particle has its own freezeout surface.

\begin{figure*}
\centering
 \includegraphics[width=0.45\textwidth]{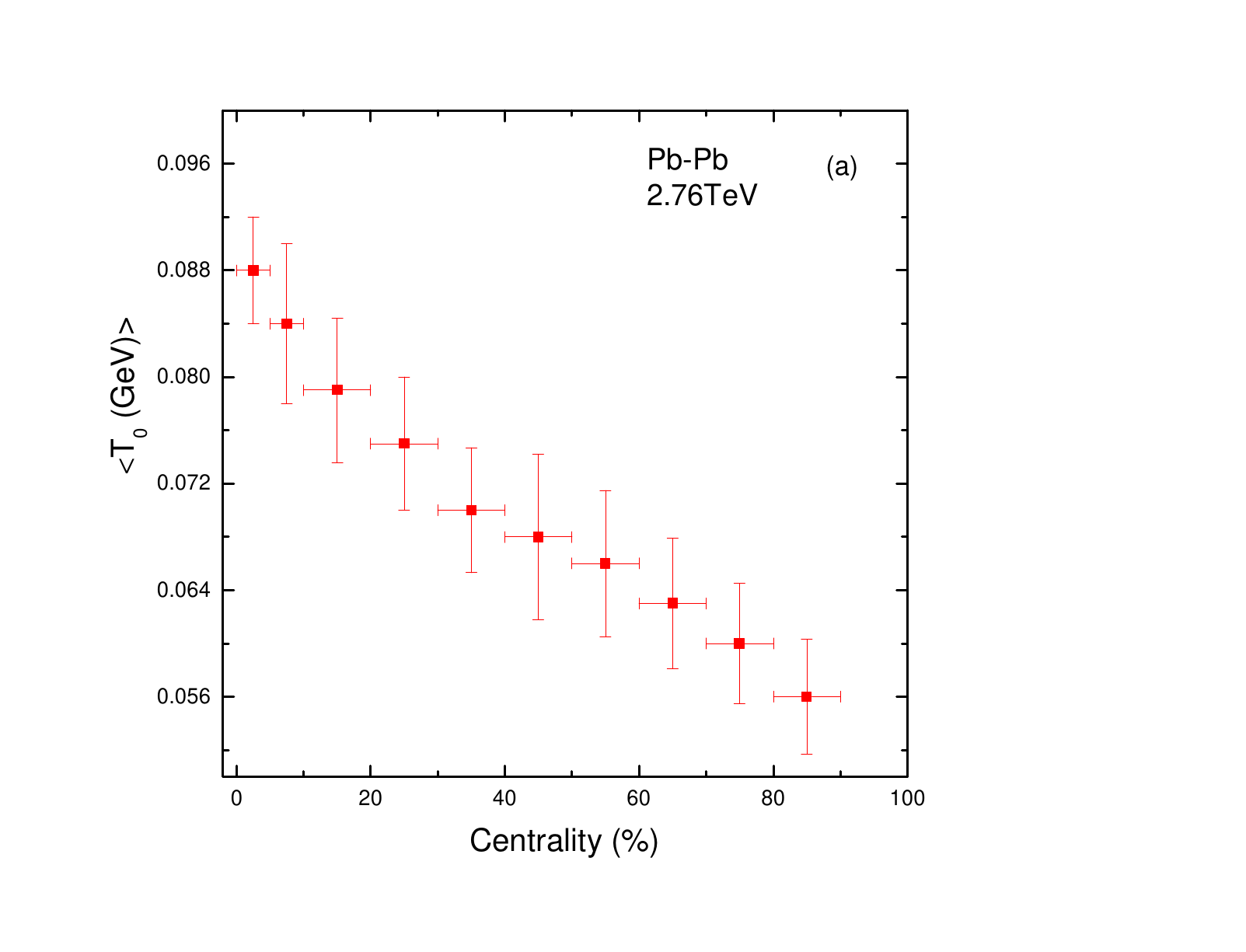}
 \includegraphics[width=0.45\textwidth]{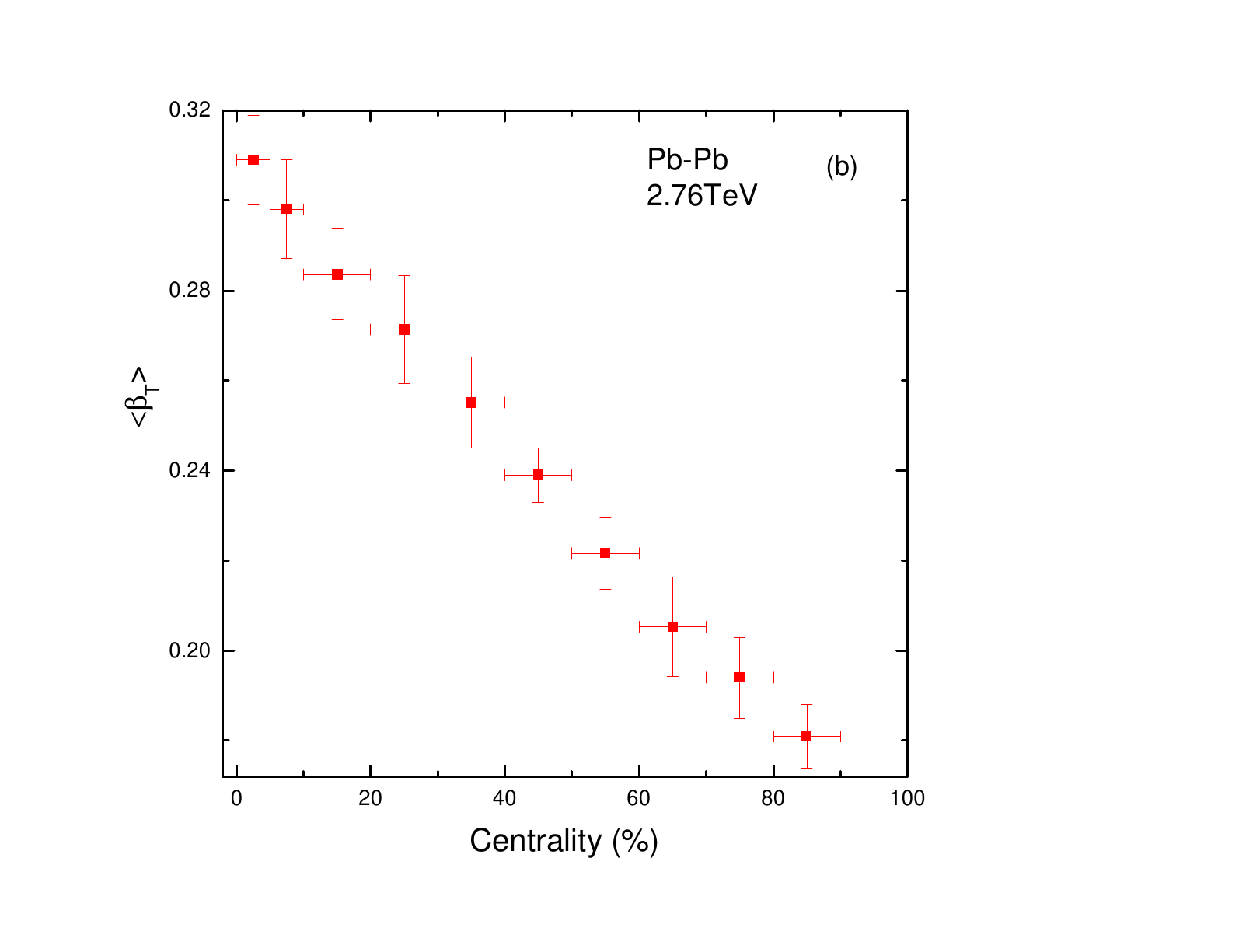}
 \includegraphics[width=0.47\textwidth]{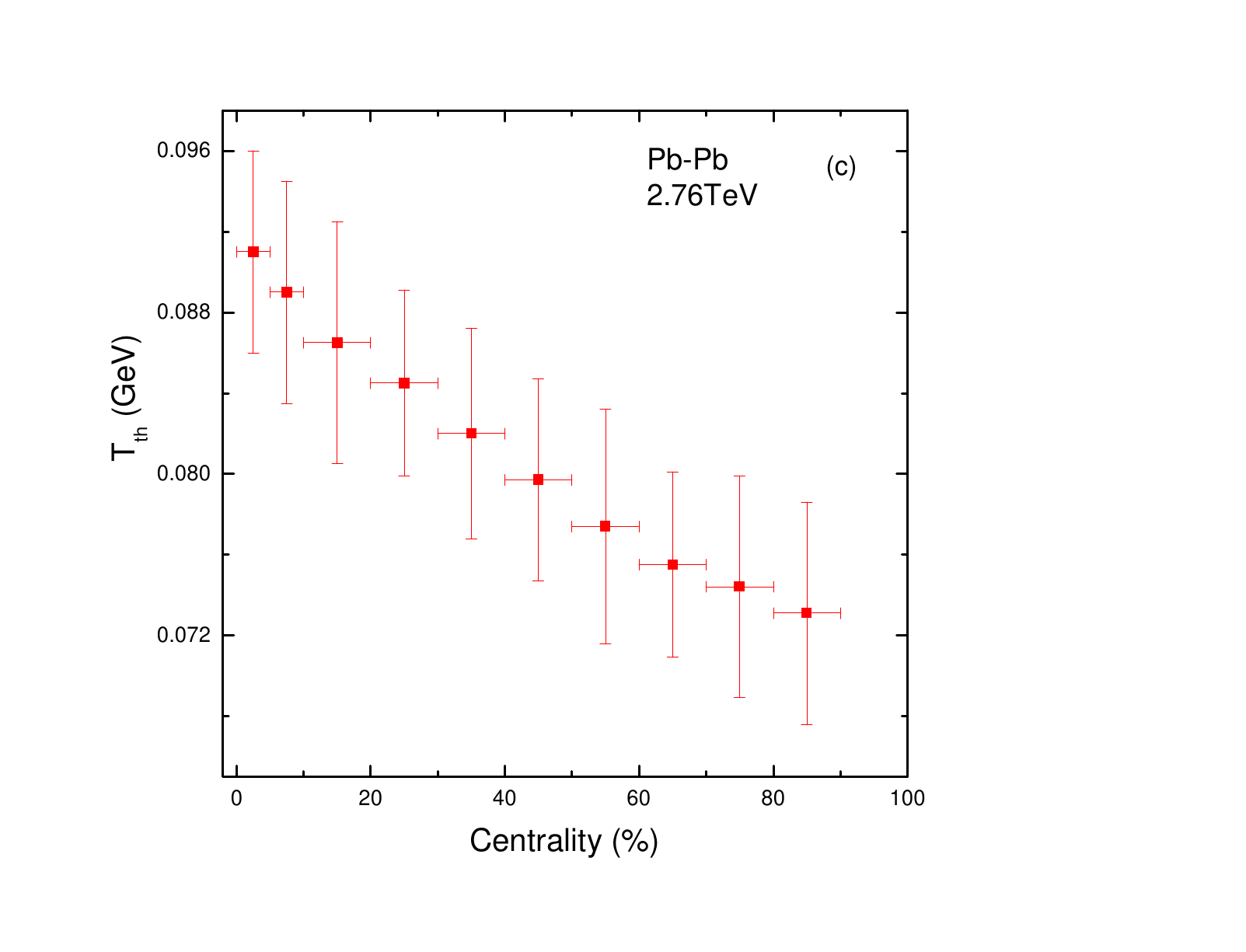}
\caption {Effects of centrality and particle mass on (a) the mean
  kinetic freezeout temperature, $\langle T_0 \rangle$, (b) the mean transverse
  flow velocity, $\langle \beta_T \rangle$, and (c) the thermal temperature, $T_{\rm th}$.}
\label{fig3}
\end{figure*}
 Figure \ref{fig3} is similar to Figure \ref{fig2}, but it shows the
result for the mean kinetic freezeout temperature $\langle T_0\rangle$,
the mean transverse flow velocity $\langle \beta_T\rangle$ and
the thermal temperature $T_{\rm th}$. We see that the qualitative
behavior of these terms with respect to centrality
is similar to $T_{\rm eff}$.
All these three quantities have a
declining trend from head-on to periphery collisions.
$\langle T_0\rangle$, in panel (a), decreases towards periphery
for the same reason as $T_{\rm eff}$. In fact, $\langle T_0\rangle$
and $T_{\rm eff}$ occur simultaneously or the latter occurs a little
earlier. Therefore, $\langle T_0\rangle$ follows
the trend of $T_{\rm eff}$. $\langle \beta_T\rangle$, in panel (b), is
maximal in head-on collisions and declines in collisions where
the impact parameter is larger. The cause of this is the
pressure gradient. In central collisions, the system has an
enormous amount of energy that results in a large pressure
gradient after the explosion, so the particles leave very
fast, which implies larger $\langle \beta_T\rangle$ values.

Our results disagree with \cite{ALICE:2013mez,STAR:2017sal,ALICE:2013wgn,ALICE:2019hno}, where $T_0$ is larger in peripheral collisions.
  However, Refs.\ \cite{Lao:2021wub,Lao:2017dtr,Waqas:2022nae} are
  in agreement with our work, where the head-on collision corresponds
  to larger $T_0$ and $\beta_T$, in agreement with a higher degree of excitation
  in the system and a quick expansion. It should be noted that Refs.\
  \cite{ALICE:2013mez,STAR:2017sal,ALICE:2013wgn,ALICE:2019hno} used
  a simultaneous fit. However, the trend of the parameters does not
  change whether the fit is individual or simultaneous, as shown in
  our previous studies \cite{Waqas:2021nku,Waqas:2018tkk}. The
  trend of $T_0$ might change if a different model is used.
  Even when analyzing the same data with the same model, but with a
  different method of parameter extraction, the results could be different
  \cite{Waqas:2019mjp}.
 
 In panel (c), the values of thermal temperature --- obtained from the
flow correction of the spectral temperatures --- are displayed.
The flow correction formula reads
 \begin{eqnarray}
\label{eq1}
T_{\rm th}=T+\sqrt \frac{1+\beta_T}{1-\beta_T},
\end{eqnarray}
We observe that $T_{\rm th}$ decreases monotonously from
head-on to peripheral collisions.

\begin{figure*}
\centering
 \includegraphics[width=0.35\textwidth]{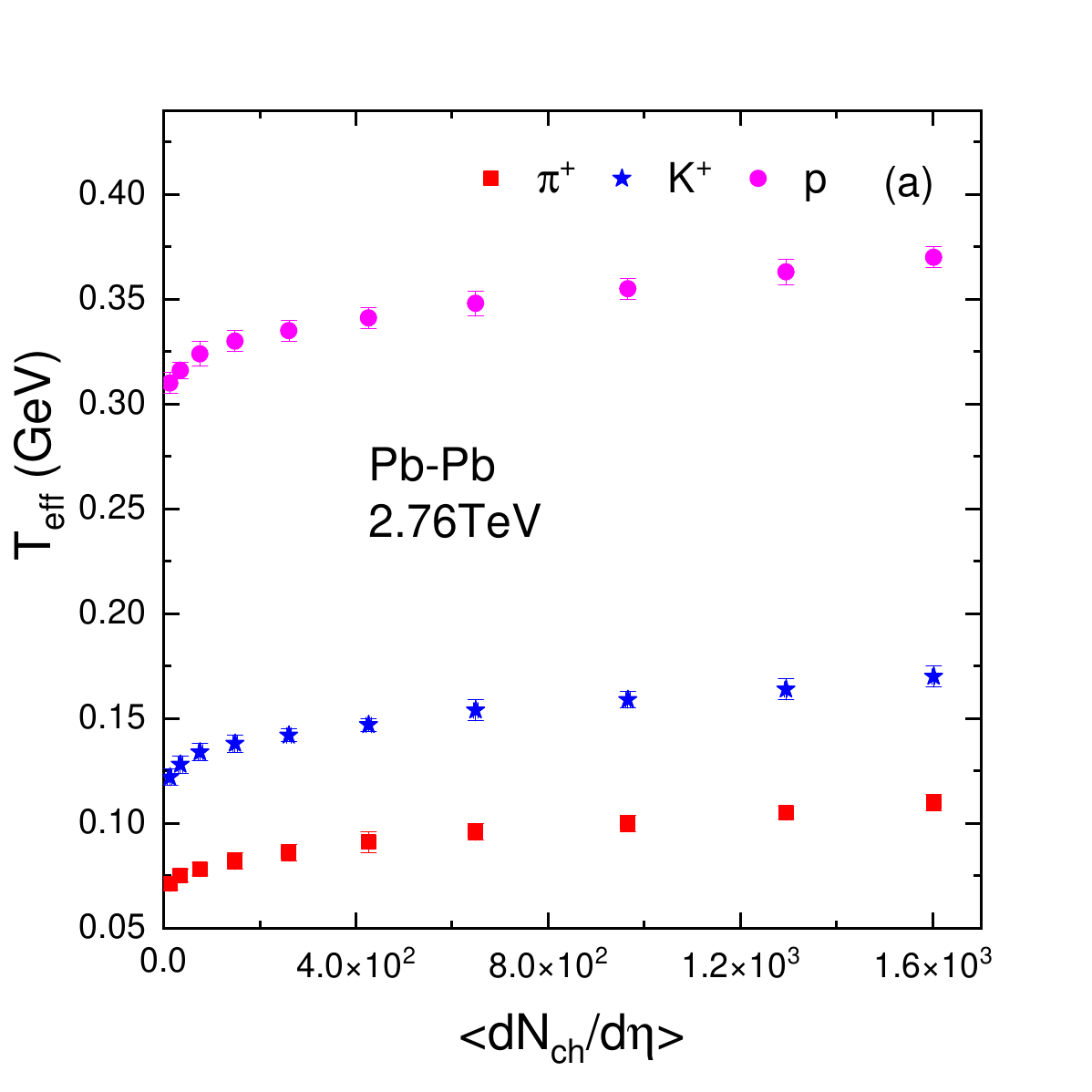}
 \includegraphics[width=0.35\textwidth]{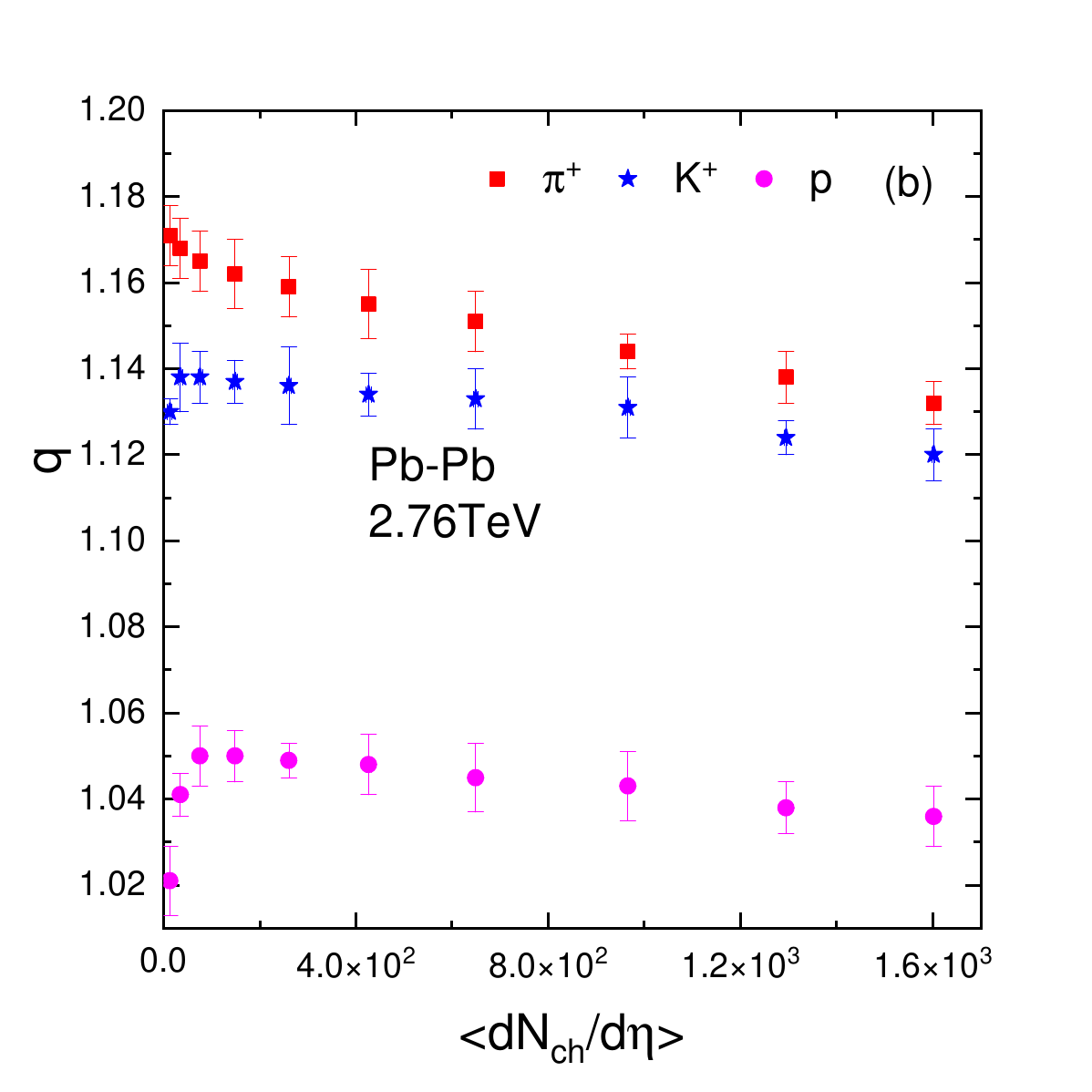}
 \includegraphics[width=0.35\textwidth]{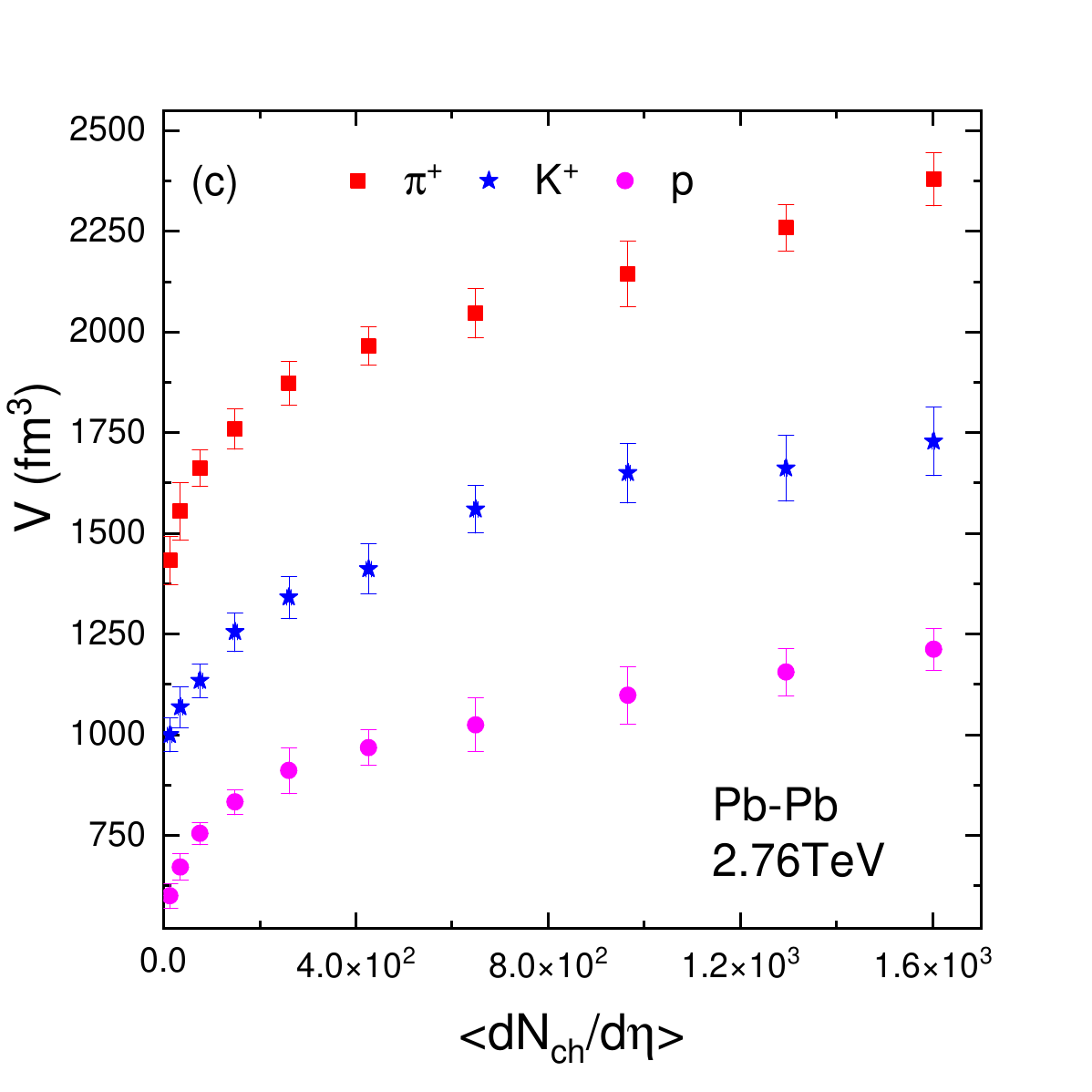}
 \includegraphics[width=0.35\textwidth]{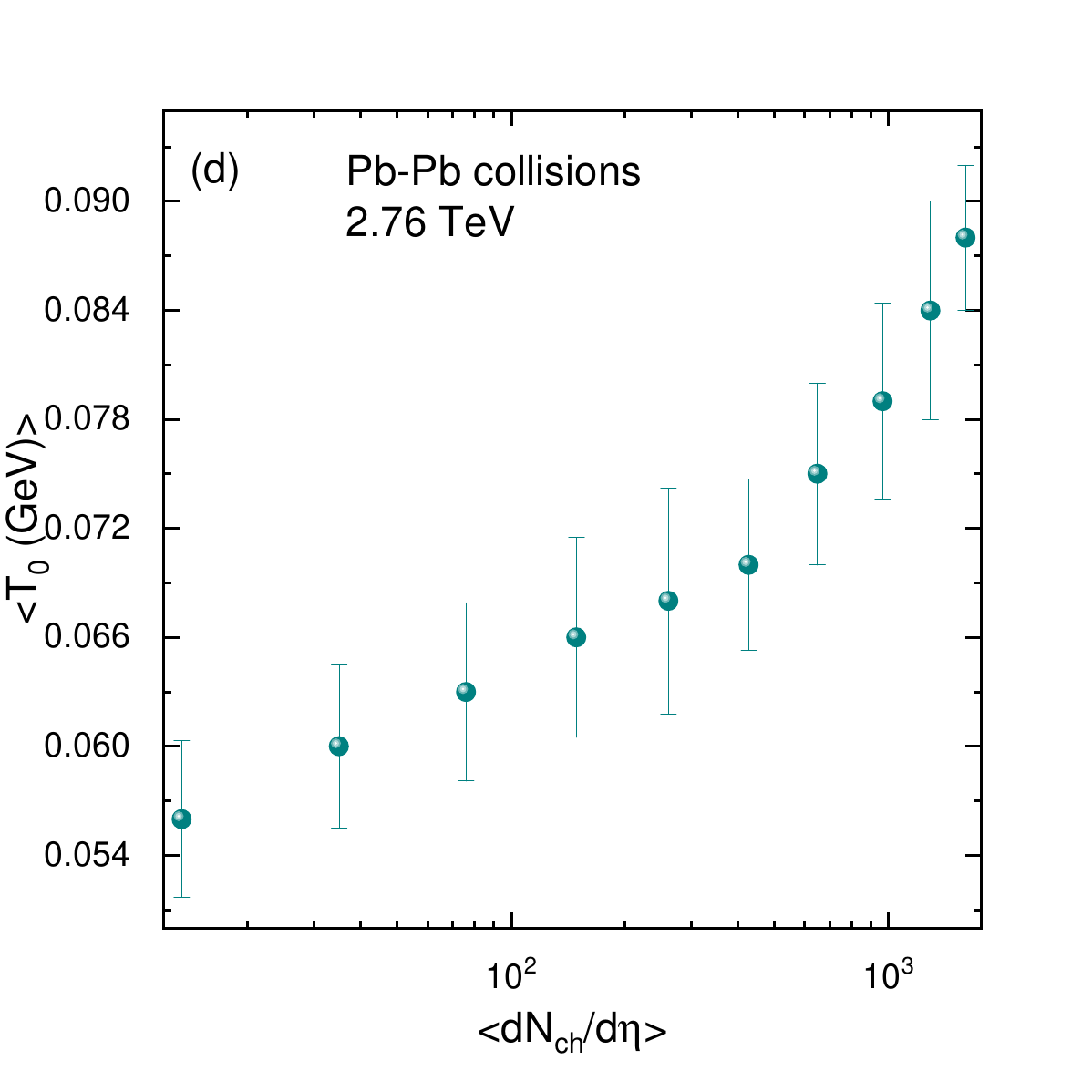}
 \includegraphics[width=0.35\textwidth]{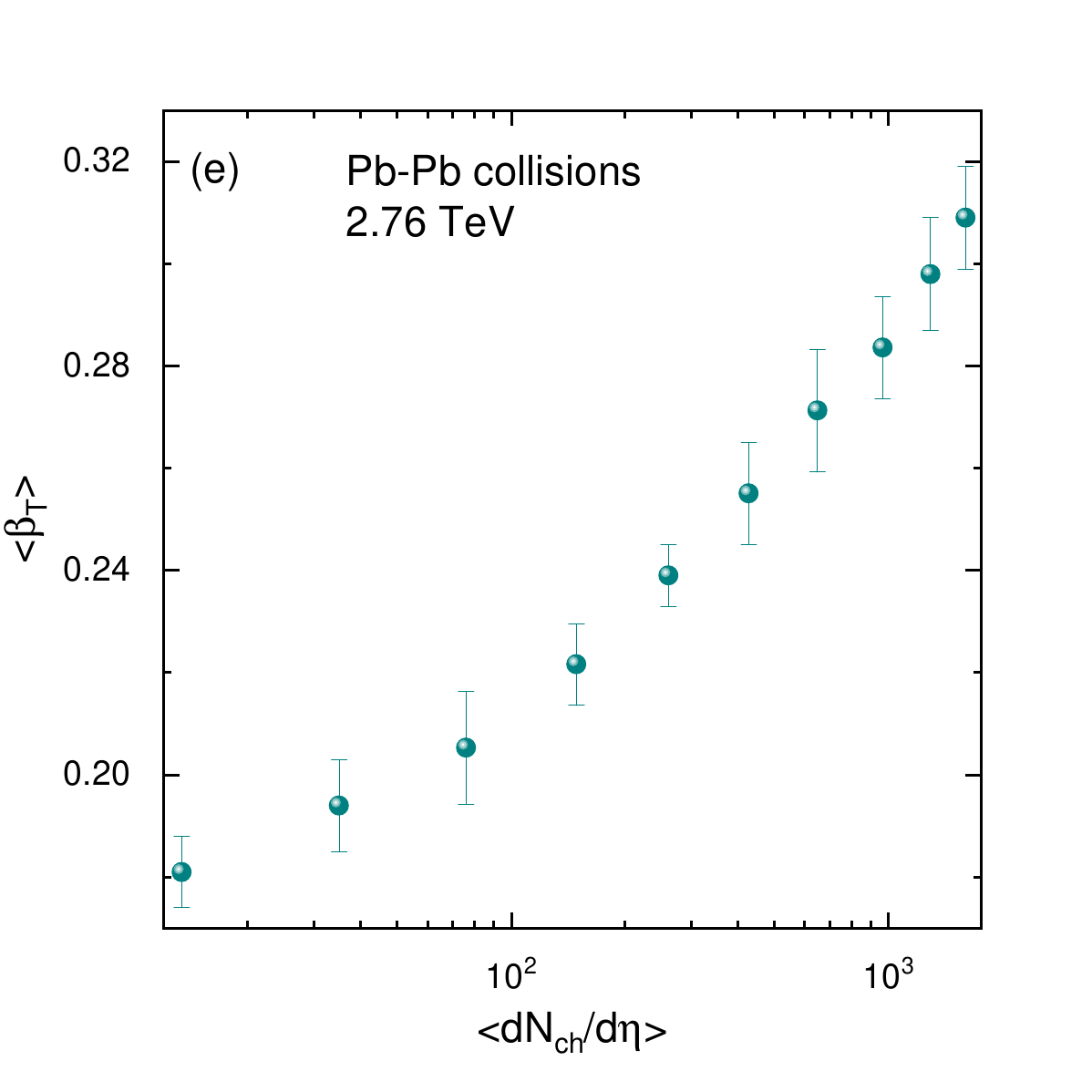}
 \includegraphics[width=0.35\textwidth]{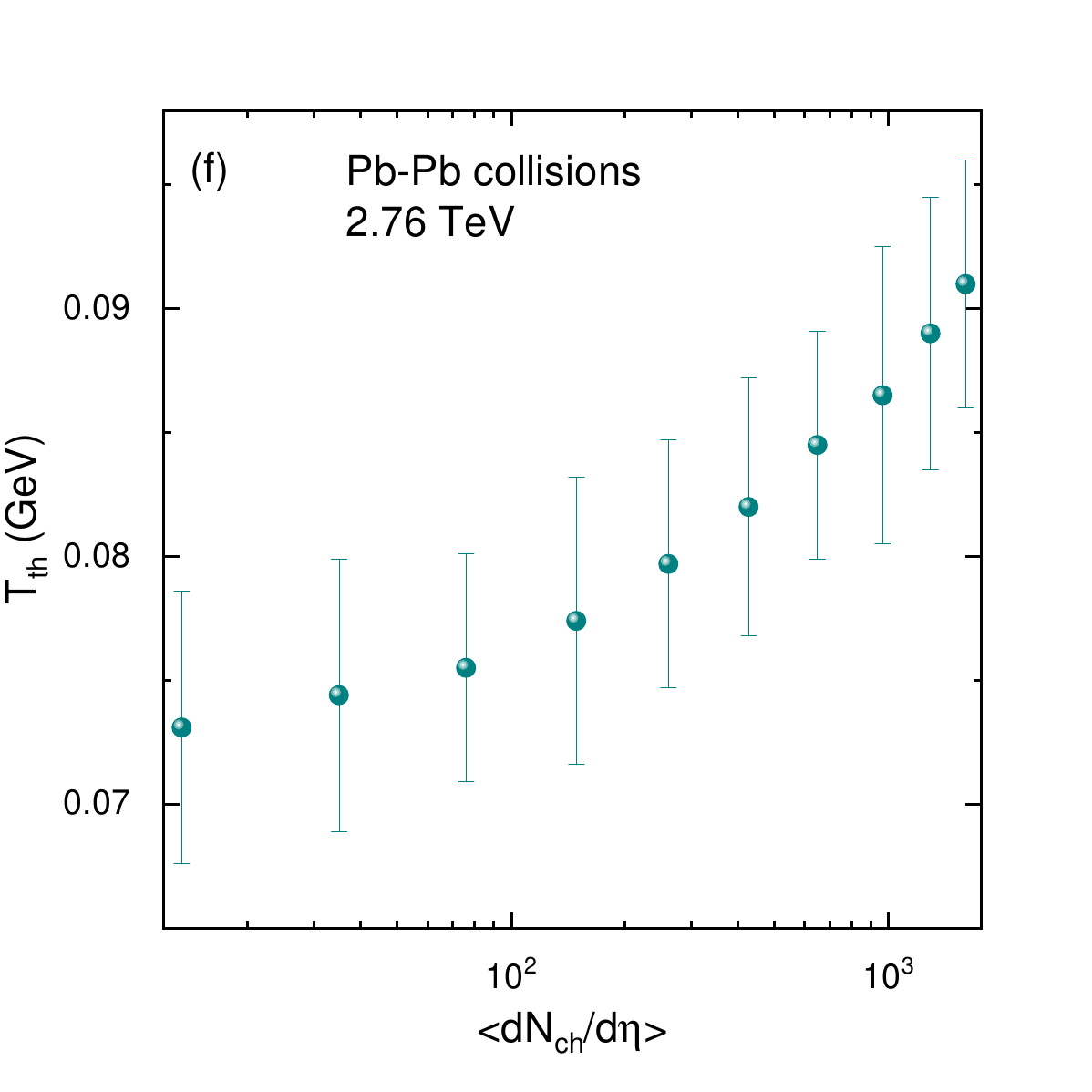}
\caption{The dependence of various quantities on
  the multiplicity of charged particles, $N_{\rm ch}$, per unit
  of pseudo-rapidity, \(\langle dN_{\rm ch}/d\eta\rangle\),
  and on $m$. Panels (a) to (c) display
  the behavior of effective temperature $T$, the non-extensitivity
  parameter $q$ and kinetic freezeout volume $V$, respectively,
  while panels (d) to (f) show the behavior of the average kinetic
  freezeout temperature $\langle T_0 \rangle$, the average transverse flow
  velocity $\langle \beta_T \rangle$ and the thermal temperature $T_{\rm th}$,
  respectively, depending on \(\langle dN_{\rm ch}/d\eta\rangle\).
  The data of \(\langle dN_{\rm ch}/d\eta\rangle\) is taken from Ref.\
  \cite{ALICE:2013mez}.}
\label{fig4}
\end{figure*}

Figure \ref{fig4} shows results for $T$, $q$, $V$,
$\langle T_0\rangle$, $\langle \beta_T\rangle$, and $T_{\rm th}$,
depending on the multiplicity of charged particles per unit
pseudorapidity, $\langle dN_{\rm ch}/d\eta \rangle$.
The results for $T$, $q$, $V$,
$\langle T_0\rangle$, $\langle \beta_T\rangle$, and $T_{\rm th}$
are shown in panels (a), (b), (c), (d), (e), and (f), respectively.
There is a direct relation between $T$ and \(\langle
dN_{\rm ch}/d\eta\rangle\). $T$ rises a little with increasing
\(\langle dN_{\rm ch}/d\eta\rangle\). $\langle T_0\rangle$
and $T_{\rm th}$ rise as well with the increase of
\(\langle dN_{\rm ch}/d\eta\rangle\).

The observed positive correlation implies that better
thermalization inside the collision system may be associated
with larger \(\langle dN_{\rm ch}/d\eta\rangle\) values.
We know that \(\langle dN_{\rm ch}/d\eta\rangle\) indicates
the intensity of the collision by measuring the density of
particles created within a particular pseudorapidity range.
A higher energy density in the system is suggested by an
increase in the multiplicity of charged particles. An environment
becomes hotter when a collision generates more charged particles
because more energy is delivered to the system. As a result, the
system conserves more thermal energy before ceasing to interact
and cooling, leading to a higher kinetic freezeout temperature.
In other words, we can say that interactions between charged
particles occur more frequently in a system with more charged
particles, which delays the freezeout process. Consequently, the
system can sustain a higher temperature until kinetic freezeout
takes place, because it stays thermalized for a longer period
before decoupling.

Furthermore, we see that $p$ has the highest, while $\pi^+$ has
the lowest temperature at fixed \(\langle dN_{\rm ch}/d\eta\rangle\).
The reason is that the system experiences
considerable collective expansion as a result of the high
pressure gradients in the dense and hot medium after the collision
of heavy ions. Particles with large mass tend to pick up more momentum
due to this radial flow. Consequently, this collective expansion
gives heavier particles more kinetic energy. Since heavier
particles have more momentum, they appear to have a higher
kinetic freezeout temperature than lighter particles as the
system cools and freezes out (i.e., particle interactions become
too rare to appreciably alter their momenta).
The momentum distribution of heavier particles
indicates a higher temperature (effective or kinetic freezeout
temperature) because collective expansion (radial flow)
contributes more to their momentum. This suggests that the
mass-dependent influence of the radial flow causes the kinetic
characteristics (momentum spectra) to appear differently for
different particle species at freezeout, but not that the
system's true thermal temperature $T_{\rm th}$ varies. 

Similarly, the
correlation between \(\langle dN_{\rm ch}/d\eta\rangle\) and $V$
turns out to be positive. As we see in Figure \ref{fig2}, with respect
to mass dependence, $V$ in panel (c) has an inverse mass dependence
to that of $T$. Hence $V$ is larger for lighter particles
at the same value of \(\langle dN_{\rm ch}/d\eta\rangle\).
Furthermore, $V$ increases with $\langle dN_{\rm ch}/d\eta\rangle$.
Since the system expands as
it evolves following the collision, this results in greater
multiplicity, indicating higher pressure and energy in
the system, which cause a larger expansion prior to the kinetic
freezeout stage. The freezeout volume is larger in the system
with higher multiplicity because it has expanded more by the time
of kinetic freezeout. In other words, when the system approaches
the point at which kinetic interactions end, the more particles
it produces, the larger the volume it fills.

Like $T$ and $V$, $\langle \beta_T\rangle$ in panel (e)
of Figure \ref{fig4} also grows
with increasing \(\langle dN_{\rm ch}/d\eta\rangle\), because large
\(\langle dN_{\rm ch}/d\eta\rangle\) corresponds to higher energy
deposition in the system which results in a large pressure gradient,
and hence larger $\langle \beta_T\rangle$. 

However, the non-extensitivity parameter $q$ shows a strange
behavior --- distinct from the other
parameters --- with respect to \(\langle dN_{\rm ch}/d\eta\rangle\),
see panel (b). For $\pi^+$ it continuously decreases from lower to
higher \(\langle dN_{\rm ch}/d\eta\rangle\), whereas for $K^+$ and $p$
it initially increases and then remains nearly invariant in
\(\langle dN_{\rm ch}/d\eta\rangle\). At low multiplicity, the
system is farther from equilibrium, leading to a higher $q$.
As \(\langle dN_{\rm ch}/d\eta\rangle\) increases, the collisions
become more frequent, and the system moves toward thermal
equilibrium, initially increasing the correlations among
particles. Hence the system approaches equilibrium as multiplicity
rises, and the degree of non-extensitivity marginally declines.
The decrease of $q$ at high multiplicities suggest that the $\pi^+$
approach a more thermalized state in these dense
collision environments. $K^+$ and $p$ have a slightly different
production mechanism compared to $\pi^+$. At low multiplicity,
the production of $K^+$ is more sensitive to fluctuations and
correlations, leading to an initial increase in $q$. As the
multiplicity increases, the $K^+$ and $p$ production stabilizes,
leading to a situation where $q$ remains almost constant. This
suggests that the $K^+$ and $p$ ensembles reach a relatively stable
non-equilibrium state even as \(\langle dN_{\rm ch}/d\eta\rangle\)
increases further.

\begin{figure*}
\centering
 \includegraphics[width=0.49\textwidth]{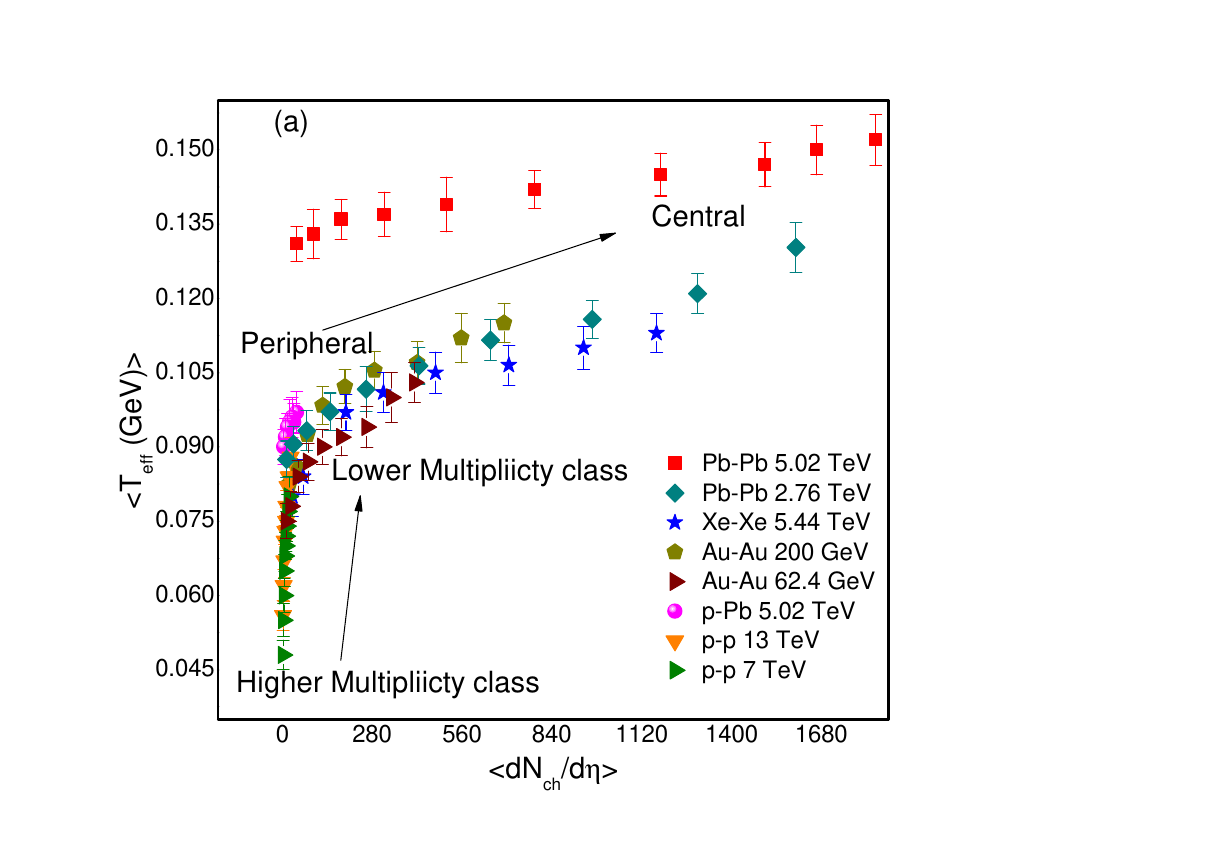}
 \includegraphics[width=0.49\textwidth]{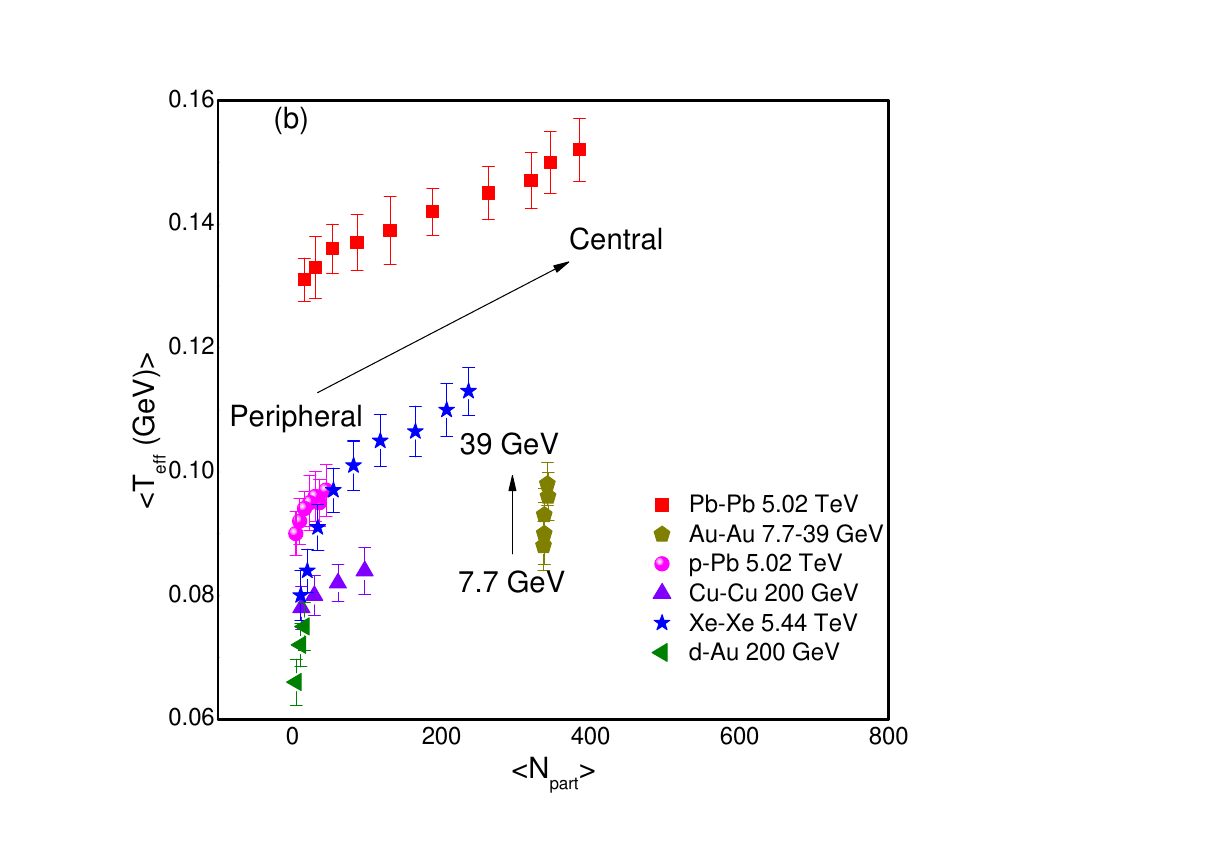}
 \caption{The results of $\langle T_{\rm eff} \rangle$ versus
   (a) \(\langle dN_{\rm ch}/d\eta\rangle\),
   and (b) \(\langle N_{\rm part}\rangle\).}
\label{fig5}
\end{figure*}

It is important to investigate the dependencies
of $T$ and $q$ on \(\langle N_{\rm part} \rangle\) and on
\(\langle dN_{\rm ch}/d\eta\rangle\) for the above collision systems
at the LHC and RHIC. We calculated \(\langle T\rangle\) by the
weighted average of $T$ for $\pi^+$, $K^+$ and $p$ produced in
Pb-Pb collisions at 2.76 TeV. We further extracted $\la T\ra$ for the
particles in different collision systems and energies at RHIC
and LHC, and calculated \(\langle T_{\rm eff}\rangle\) by taking their
weighted average. Figure \ref{fig5} shows \(\langle T_{\rm eff}\rangle\)
as a function of \(\langle dN_{\rm ch}/d\eta\rangle\), and of
\(\langle N_{\rm part}\rangle\). The figure shows these dependencies
over different systems and energies. It provides
a clear comparison of the correlation of \(\langle T_{\rm eff}\rangle\)
and \(\langle dN_{\rm ch}/d\eta\rangle\), and \(\langle T_{\rm eff}\rangle\)
and \(\langle N_{\rm part}\rangle\) in different collision systems
and at various energies.

In panel (a), \(\langle T_{\rm eff}\rangle\) is presented as a
function of \(\langle dN_{\rm ch}/d\eta\rangle\), while panel (b)
shows \(\langle T_{\rm eff}\rangle\) as a function of
\(\langle N_{\rm part}\rangle\). Different symbols in Figure \ref{fig5}
display different collision systems at different energies.
Panel (a) shows \(\la T_{\rm eff}\ra \) as a function
of \(\langle dN_{\rm ch}/d\eta\rangle\) in different systems, namely
Pb-Pb collisions at 2.76 and 5.02 TeV, Xe-Xe collisions
at 5.44 TeV, Au-Au collisions at 200 and 62.4 GeV, p-Pb
collisions at 5.02 TeV, and p-p collisions at 7 and 13 TeV.
Panel (b) displays \(\langle T_{\rm eff}\rangle\)
as a function of \(\langle N_{\rm part}\rangle\) for
Pb-Pb collisions at 5.02 TeV, Au-Au collisions from
7.7 to 39 GeV, p-Pb collisions at 5.02 TeV, Cu-Cu collisions at
200 GeV, Xe-Xe collisions at 5.44 TeV, and d-Au collisions at
200 GeV. The data of \(\langle dN_{\rm ch}/d\eta\rangle\)
is taken from Refs.\ \cite{ALICE:2013mez,ALICE:2015juo,STAR:2008med,ALICE:2018pal,ALICE:2020nkc,ALICE:2013wgn,ALICE:2018hza}, while the data of
\(\langle N_{\rm part}\rangle\) is taken from Refs.\ \cite{STAR:2017sal,BRAHMS:2016klg,ALICE:2013wgn,ALICE:2018hza,STAR:2007poe,ALICE:2015juo}.

In panel (a), there are distinct trends in the correlation between
the effective temperature \(\langle T_{\rm eff}\rangle\) and the
charged-particle multiplicity density \(\langle dN_{\rm ch}/d\eta\rangle\)
for various collision systems and energies, which captures the
fundamental dynamics of particle generation. When the energy
in Pb-Pb collisions is increased from 2.76 TeV to 5.02 TeV,
the effective temperature rises for similar values of
\(\langle dN_{\rm ch}/d\eta\rangle\). At \(\langle dN_{\rm ch}/d\eta\rangle\)
$\approx 1500$, for example, \(\langle T_{\rm eff}\rangle\) increases
from roughly 0.121 (at 2.76 TeV) to roughly 0.15 (at 5.02 TeV).
This implies that, most likely as a result of higher initial energy
density, collisions with more energy produce a hotter medium.
For the same centrality bin, Au-Au collisions show a qualitatively
similar pattern, with $\la T_{\rm eff} \ra$ at 200 GeV
consistently larger than at 62.4 GeV. 

In the \(\langle dN_{\rm ch}/d\eta\rangle\) dependency of
\(\langle T_{\rm eff}\rangle\), the system size is a significant
factor. Multiplicity and temperature are considerably more
correlated in heavy-ion systems (Pb-Pb, Xe-Xe, Au-Au) than in
smaller systems (p-Pb, p-p). While \(\langle T_{\rm eff}\rangle\)
fluctuates between 0.131 and 0.152 in central Pb-Pb collisions
at 5.02 TeV across a wide range of \(\langle dN_{\rm ch}/d\eta\rangle\),
45–1850, it only varies between 0.056 and 0.088 in p-p
collisions at 13 TeV over a substantially smaller
\(\langle dN_{\rm ch}/d\eta\rangle\) interval (2.55–26.02).
This discrepancy suggests that the sensitivity of
\(\langle T_{\rm eff}\rangle\) to the multiplicity is increased
by the substantial collective effects that massive collision
systems undergo, such as hydrodynamic flow and quark-gluon
plasma production. Furthermore, we find that
\(\langle T_{\rm eff}\rangle\) is typically lower in Xe-Xe
collisions at similar multiplicities when comparing Xe-Xe
at 5.44 TeV with Pb-Pb collisions at 5.02 TeV. This is probably
because there are fewer participating nucleons, and the system
size is smaller. In contrast, p-Pb and p-p collisions exhibit a
significantly reduced reliance of \(\langle T_{\rm eff}\rangle\)
on \(\langle dN_{\rm ch}/d\eta\rangle\), indicating that mechanisms
such as string fragmentation or mini-jet processes, rather than
thermalization, dominate the particle creation in these systems.
With values of around 0.14–0.15~GeV for Pb-Pb at 5.02~TeV, and
$\sim 0.115$~GeV for Au-Au at 200 GeV, \(\langle T_{\rm eff}\rangle\)
appears to approach a saturation-like behavior at very high
multiplicities (central heavy-ion collisions).

On the other hand, \(\langle T_{\rm eff}\rangle\) drastically decreases
in low-multiplicity p-p and p-Pb collisions, with values as low
as $\sim 0.05–0.07$~GeV, reinforcing the scenario that small systems
do not have the powerful collective effects found in heavy-ion
collisions. With heavy-ion collisions displaying a unique
thermal-like response to multiplicity, and small systems displaying
a weaker, potentially non-thermal dependence, the findings
collectively indicate that the effective temperature is highly
impacted by both collision energy and system size. This is
consistent with the predictions of QCD thermodynamics, which
states that perturbative processes still dominate smaller
systems, whereas the quark-gluon plasma behavior is more likely
to show up in larger and more energetic systems. 

We repeat that panel (b) shows \(\langle T_{\rm eff}\rangle\) as a
function of \(\langle N_{\rm part}\rangle\). There are fundamental
shifts in particle formation mechanisms among the collision systems,
as illustrated by the dependence of the effective temperature on
the participant number. One can see a substantial increase of
\(\langle T_{\rm eff}\rangle\) with \(\langle N_{\rm part}\rangle\)
in Pb-Pb collisions at 5.02 TeV, from 0.131~GeV at
\(\langle N_{\rm part}\rangle\) $\approx$ 15 to 0.152~GeV at
\(\langle N_{\rm part}\rangle\) $\approx$ 385. This remarkable
expansion suggests that large systems at LHC energy have strong
collective effects. This pattern is in sharp contrast to Au-Au
collisions at lower energy (7.7-39 GeV), where
\(\langle T_{\rm eff}\rangle\) is nearly constant ($0.088-0.098$~GeV),
even if the participant numbers are identical. This comparison
demonstrates the significance of the collision energy in the
process of thermalization.

The behavior of intermediate systems is transitional.
At 5.44 TeV, Xe-Xe collisions indicate a considerable
\(\langle N_{\rm part}\rangle\) dependence, with
\(\langle T_{\rm eff}\rangle\) growing from 0.08 to 0.113~GeV over the
range of measured participant numbers. Even at the same energies,
there are noticeable system size effects, since the temperatures
in Xe-Xe are roughly $20\%$ lower than in Pb-Pb at comparable
\(\langle N_{\rm part}\rangle\) $\approx 150-200$. Even more
suppressed temperatures (0.078-0.084~GeV) have been observed in
the smaller Cu-Cu system at 200 GeV, emphasizing the significance
of the system volume for thermalization. The characteristics of
small collision systems are basically different. The
\(\langle T_{\rm eff}\rangle\) versus \(\langle N_{\rm part}\rangle\)
distributions are almost flat for both p-Pb at 5.02 TeV and d-Au
at 200 GeV, with p-Pb retaining \(\langle T_{\rm eff}\rangle\)
$\approx 0.09-0.097$~GeV, and d-Au ranging only from 0.066 to 0.075~GeV.
Rather than collective effects, this modest dependence suggests that
non-thermal processes dominate the particle formation in these small
systems. Remarkably, the p-Pb and Xe-Xe peripheral systems
reach comparable temperatures in spite of their disparate sizes,
which may suggest a shared limiting mechanism for small-scale
particle generation. 

From this comparison, multiple significant trends can be seen.
First, at high energies ($>$10 GeV/nucleon), a strong
\(\langle N_{\rm part}\rangle\) dependency of \(\langle T_{\rm eff}\rangle\)
appears only in very large systems (usually $>50$ participants).
Likewise, at fixed \(\langle N_{\rm part}\rangle\), the temperature
rises drastically with the collision energy, indicating the critical
importance of the energy density in system thermalization. Third,
across several energy regimes, the system size hierarchy in temperatures
remains unaltered,
(Pb-Pb $>$ Xe-Xe $>$ Au-Au $\approx$ Cu-Cu $>$ p-Pb $>$ d-Au)
suggesting that the amount of matter involved
in the collision has a fundamental impact on the possible temperature.

\begin{figure*}
\centering
 \includegraphics[width=0.49\textwidth]{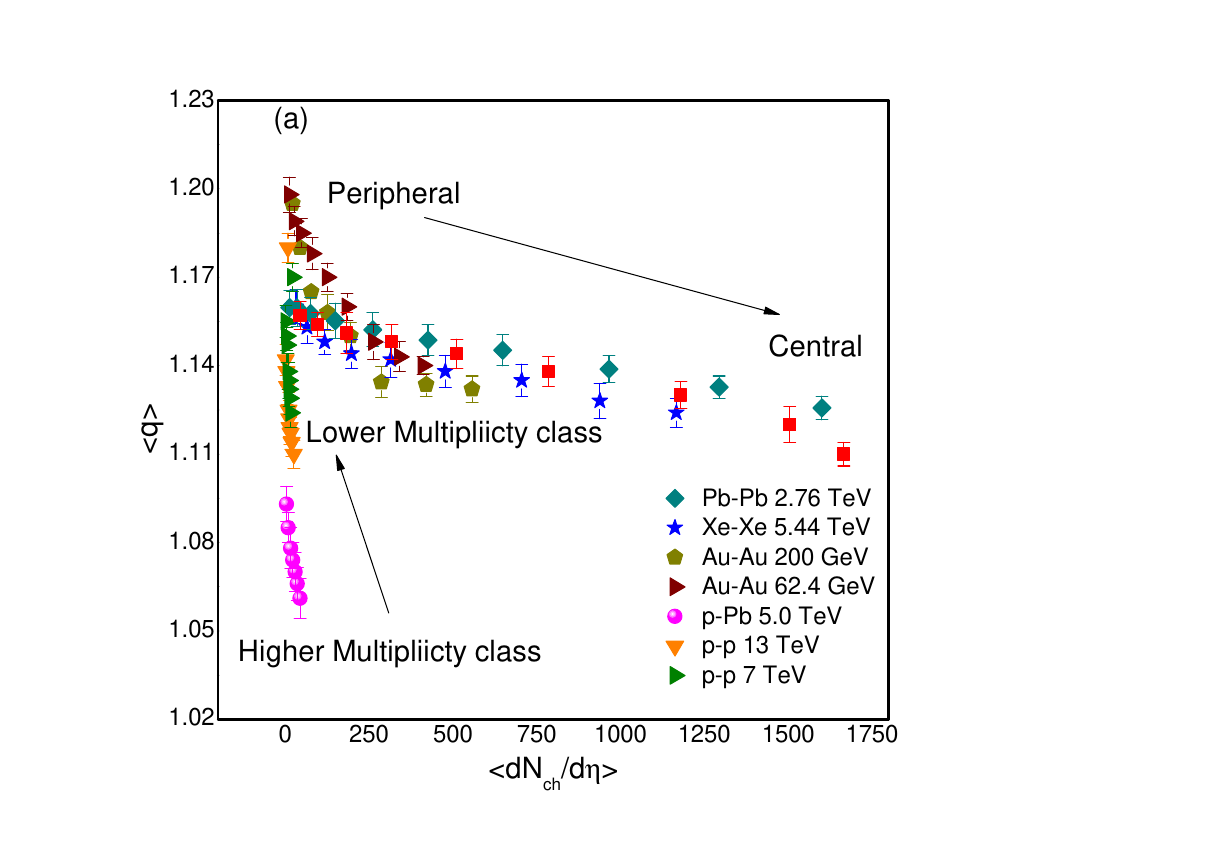}
\includegraphics[width=0.49\textwidth]{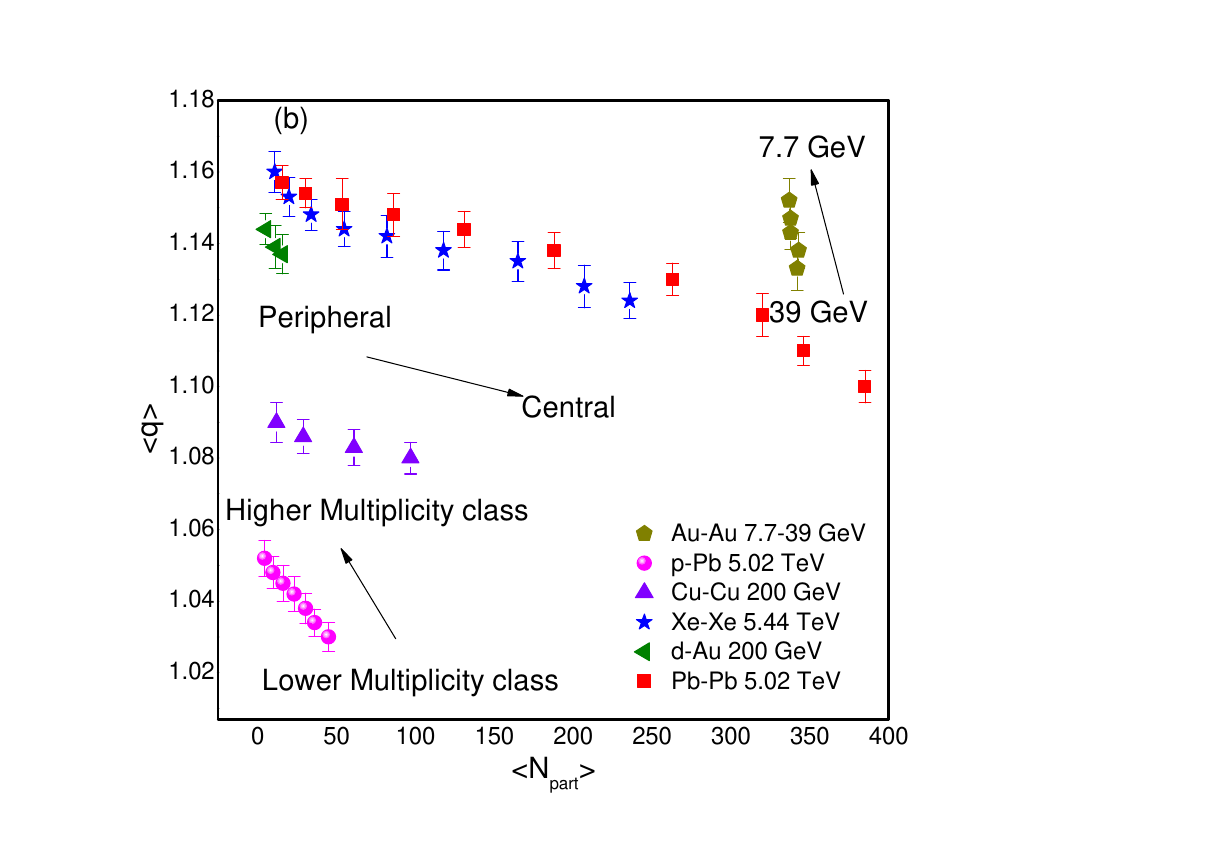}
\caption{The results of $\langle q \rangle $ versus
  (a) \(\langle dN_{\rm ch}/d\eta\rangle\), and
  (b) \(\langle N_{\rm part}\rangle\).}
\label{fig6}
\end{figure*}

Figure \ref{fig6} is similar to Figure \ref{fig5}, but it
shows the correlation of $\la q \ra$ with \(\langle dN_{\rm ch}/d\eta\rangle\)
and \(\langle N_{\rm part}\rangle\) in panel (a) and (b), respectively.
Panel (a) reveals several basic patterns in the behavior of the
parameter $\la q \ra$ in various collision systems.
The average non-extensitivity $\la q \ra$ systematically
increases with decreasing \(\langle dN_{\rm ch}/d\eta\rangle\) in
heavy-ion collisions (Pb-Pb, Xe-Xe, and Au-Au), corresponding
to more peripheral collisions. At 2.76 TeV, $\la q\ra$ increases from
1.1256 in central collisions ($\la dN_{\rm ch}/d\eta\ra
\approx 1600$) to 1.1597 in peripheral collisions
(\(\langle dN_{\rm ch}/d\eta\rangle\) $\approx 13.4$), a tendency
that is especially pronounced in Pb-Pb systems. A similar but
slightly weaker trend exists in Pb-Pb at 5.02 TeV, exhibiting
that greater collision energies may modestly reduce the non-extensive
effects. The Xe-Xe system at 5.44 TeV exhibits a drastically
different behavior, with $\la q\ra$-values dropping with
\(\langle dN_{\rm ch}/d\eta\rangle\) significantly below 1.2.
This suggests that different non-extensive thermodynamics result
from the smaller system size of Xe-Xe, as opposed to Pb-Pb,
maybe as a result of less collective effects. At RHIC energies
(200 GeV and 62.4 GeV), the Au-Au systems exhibit a more intricate
patterns. They attain notably high values ($\la q\ra > 1.3$) in central
200 GeV collisions, with $\la q\ra$ first falling and then increasing
with \(\la dN_{\rm ch}/d\eta\ra \).

The features of small systems (p-Pb, p-p) are distinct from those of
heavy-ion collisions. While p-p collisions exhibit greater variability,
including some exceptionally high $\la q\ra$-values (e.g.\ $q = 1.41$
at $\la dN_{\rm ch}/d\eta\rangle \approx 6.7$ in 7 TeV p-p), the p-Pb
system exhibits relatively low $\la q\ra$-values ($1.06-1.09$)
with a weak dependence on \(\langle dN_{\rm ch}/d\eta\rangle\).
Compared to heavy-ion systems, the p-p systems often exhibit a
less systematic \(\langle dN_{\rm ch}/d\eta\rangle\)-dependency.

The energy dependence is best studied by comparing similar systems
at different energies.
For example, with comparable
\(\langle dN_{\rm ch}/d\eta\rangle\), Pb-Pb at 5.02 TeV has
somewhat lower $\la q\ra$-values than at 2.76 TeV, and p-p at 13 TeV
has consistently lower $\la q\ra$ than p-p at 7 TeV. This suggests
that systems may approach the Boltzmann statistics ($q \to 1$)
with larger collision energy, presumably as a result of more
thermalization. When comparing collisions with similar energy,
the hierarchy of system sizes becomes evident. We find
$\la q\ra \approx 1.13-1.15$ in Pb-Pb,
$\la q\ra \approx 1.10-1.13$ in Xe-Xe,
$\la q\ra \approx 1.06-1.07$ in p-Pb, and
$\la q\ra \approx 1.11-1.14$ in p-p
collisions for $\la dN_{\rm ch}/d\eta\ra \approx 100-200$.
This ordering by system size is consistent over energy regimes
and indicates that the overall collision geometry has a significant
impact on the degree of non-extensitivity. These trends most likely
represent basic variations in the thermalization mechanisms of
various collision systems. While the approximation $q \approx 1$ in
central heavy-ion collisions demonstrates more developed thermal
characteristics, the stronger non-extensive behavior (higher $\la q\ra$)
in peripheral heavy-ion collisions may indicate less complete
thermalization. The energy dependence may result from shifting
soft-hard process balances or from quark-gluon plasma production
becoming more dominant at higher energies. For theoretical models
of non-extensive thermodynamics in high-energy collisions, the
findings offer crucial constraints. 

In addition, panel (b) demonstrates several basic trends in the
behavior of the non-extensitivity parameter $\la q\ra$ as a
function of participant number \(\langle N_{\rm part}\rangle\).
From $\la q \ra = 1.10$ at $\langle N_{\rm part}\rangle \approx 385$
to $\la q \ra = 1.157$ at $\langle N_{\rm part}\rangle \approx 15.6$,
$\la q \ra$ increases monotonically in central Pb-Pb collisions at
5.02 TeV, confirming a stronger non-extensive behavior in more peripheral
collisions. In Xe-Xe collisions at 5.44 TeV, this pattern is
qualitatively similar, although the $\la q\ra$-values are consistently
lower (1.124-1.160 over the range
$\langle N_{\rm part}\rangle \approx 236-11$),
which replicates the expected system size dependency. As
\(\langle N_{\rm part}\rangle\) grows from 337 to 342, the Au-Au
systems at RHIC energies (7.7-39 GeV) reveal the opposite trend,
with $\langle q \rangle$ decreasing from 1.152 to 1.133, which reflects
distinct thermalization dynamics at lower energies. Systems that are
smaller have distinctive characteristics. With a weak
\(\langle N_{\rm part}\rangle\) dependence, the p-Pb collisions
at 5.02 TeV keep much lower $\la q\ra$-values ($1.03-1.052$), whereas the
Cu-Cu collisions at 200 GeV display an intermediate behavior
($\la q\ra =1.08-1.09$).
It is interesting to note that d-Au collisions at 200~GeV exhibit
unusually elevated $\langle q \rangle$ values ($1.137-1.144$) that surpass
even central Pb-Pb data, which may suggest specific particle creation
processes in this asymmetric system. The energy dependence is
visible when comparing systems that are similar: In comparison to
LHC systems, Au-Au at RHIC energies has an inverse
$\la q \ra$-\(\langle N_{\rm part}\rangle\) correlation, while
Pb-Pb at 5.02~TeV displays greater $\la q\ra$-values than
Xe-Xe at 5.44~TeV for comparable \(\langle N_{\rm part}\rangle\).
This suggests an energy barrier where the \(\la N_{\rm part}\ra \)
dependency of $\la q \ra$ is reverse.

The hierarchy of system sizes is evident when comparing
the collisions Pb-Pb, p-Pb, Cu-Cu, and Xe-Xe, for instance
$\mbox{p-Pb} < \mbox{Cu-Cu} < \mbox{Xe-Xe} < \mbox{Pb-Pb}$.

\begin{figure*}
\centering
 \includegraphics[width=0.49\textwidth]{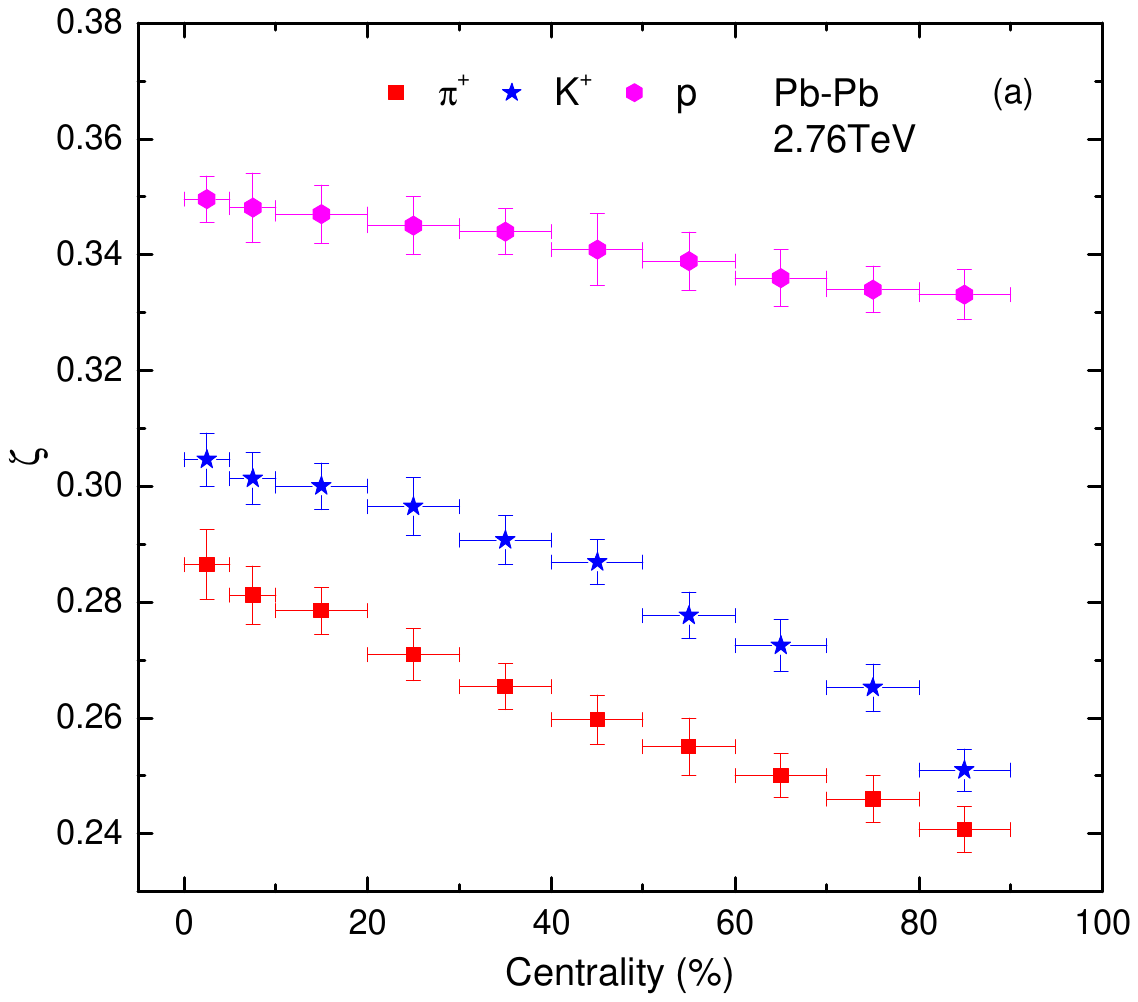}
 \includegraphics[width=0.49\textwidth]{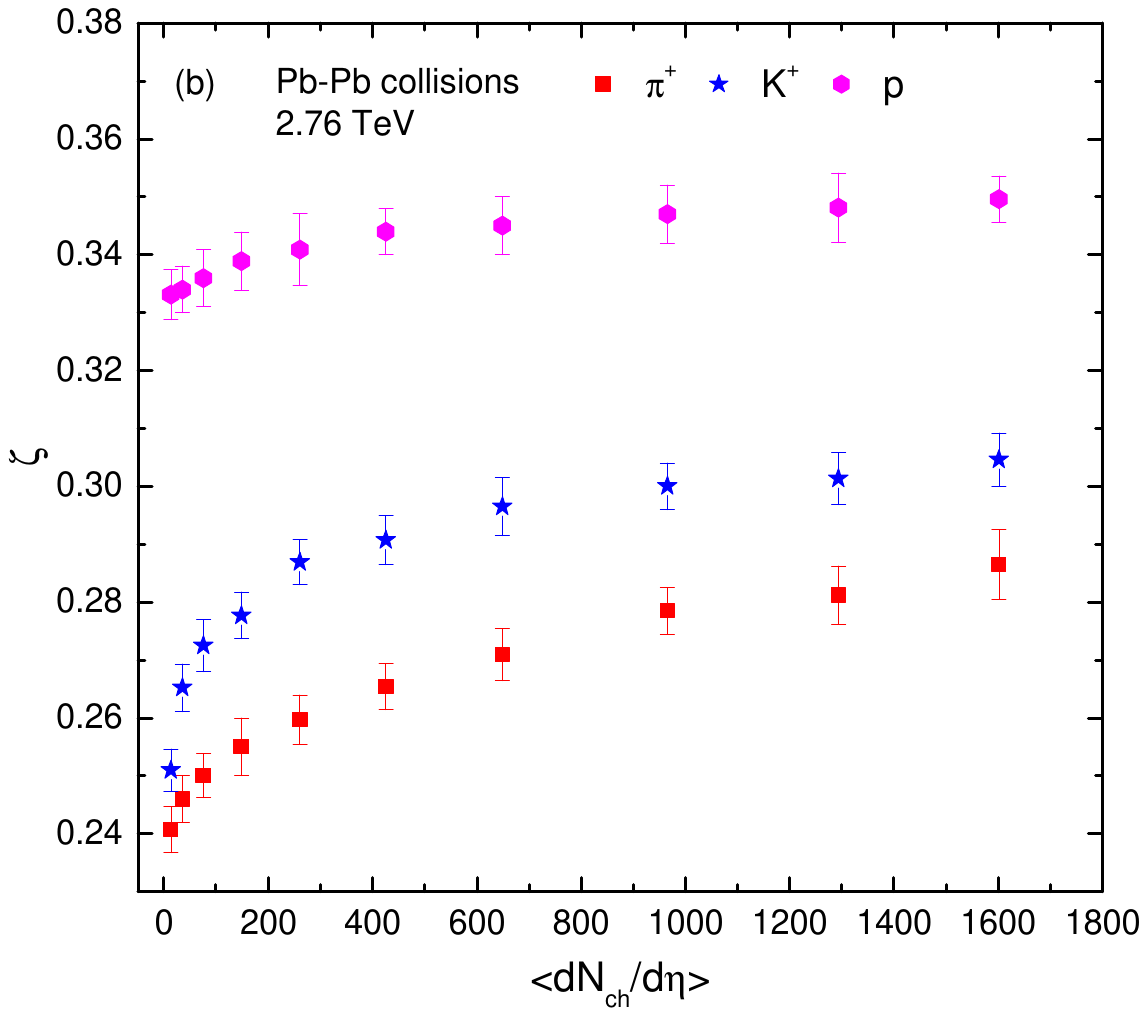}
 \caption{The result for the parameter $\zeta$ as a function of
   (a) centrality, and (b) of \(\langle dN_{\rm ch}/d\eta\rangle\).}
\label{fig7}
\end{figure*}
 The parameter $\zeta$ of Eq.\ (\ref{eq6})
is shown in Figure \ref{fig7} as a function of (a) hadron
mass and centrality, and (b) hadron mass and
\(\langle dN_{\rm ch}/d\eta\rangle\). $\zeta$ is a measure of the
system's fluctuations, including variations in energy density,
temperature and other thermodynamic parameters. The variance
of temperature, or other parameters pertaining to the dynamics
of the collision system, is usually linked to $\zeta$
in heavy ion collisions. A larger $\zeta$ value emerges in center
collisions (larger \(\langle dN_{\rm ch}/d\eta\rangle\)), because of
the increased interactions and fluctuations in the generated
medium caused by the higher density of participant nucleons.
Peripheral collisions have less oscillations because of the
lower interaction density. Thus, in Figure \ref{fig7}, as one
moves from the center to the periphery (from large to small
\(\langle dN_{\rm ch}/d\eta\rangle\)), $\zeta$ decreases.
This indicates that the lower density in peripheral collisions
(smaller \(\langle dN_{\rm ch}/d\eta\rangle\)) results in
less pronounced and smaller fluctuations, while the central
region (larger \(\langle dN_{\rm ch}/d\eta\rangle\)) has higher
energy densities and more fluctuations because of the number
of participant nucleons.
 
 In the context of heavy ion collisions
and the Tsallis distribution, the degree of correlation between
the temperature $T$ and the non-extensitivity parameter $q$ is
described by the parameter $\zeta$. In the current work, it
turns out that $\zeta$ informs us how strongly $T$ and $q-1$ are
correlated. It decreases from large to small
\(\langle dN_{\rm ch}/d\eta\rangle\), and from central to peripheral
collisions, indicating that the correlation between $T$
and $(q-1)$ weakens as the collision become less central
(smaller \(\langle dN_{\rm ch}/d\eta\rangle\)). Put more simply,
because of the greater overlap of the colliding nuclei,
central collisions (larger \(\langle dN_{\rm ch}/d\eta\rangle\))
usually include more energy and particle contact.
A more complex or turbulent system may be indicated by a larger
correlation between the temperature
and the non-extensitivity parameter $q$ as a result of this strong
interaction. As the number of peripheral collisions increases (smaller
\(\langle dN_{\rm ch}/d\eta\rangle\)), the system becomes less complex
and the temperature tends to change less because there are fewer
interactions between the colliding nuclei. Because of the system's
non-extensive nature and the reduced connection between temperature
fluctuations, $\zeta$ may decrease as a result of this reduction
in interaction. Therefore, the declining $\zeta$ trend indicates
that the non-equilibrium effects, which are represented by the
Tsallis parameter $q$, have less impact on the temperature
in peripheral collisions than in central ones. 

In addition, it is evident that $\zeta$ depends on the hadron mass.
A larger $\zeta$ value is associated with a heavier hadron. The dynamics
of particle motion and creation in high-energy physics, such as
those in heavy-ion collisions, can be used to understand this.
Even with high thermal energy, heavier particles in thermalized
systems tend to have smaller velocities because their mass
restricts their mobility. This can lead to a larger variance,
denoted by $\zeta$, and higher volatility in energy
distributions. On the other hand, lighter particles can move more freely
because their lower mass results in less fluctuations and a smaller $\zeta$. 

Finally, due to the additional energy
required for their mass, heavier particles in collisions
require more energy to be created, which may also contribute to
broader energy distributions. The broader energy distributions for heavier particles refer to their total energy, influenced by their mass and thermal fluctuations. Because they require less
energy to produce, lighter particles may exhibit more constrained
distributions. Consequently, the primary cause is that heavier
particles respond differently to thermal fluctuations and energy
distributions because of their greater mass, which results in a
larger $\zeta$ value than lighter particles. 
 
\section{Conclusions}
We matched the $p_T$ spectra of the three dominant types of positively
charged hadrons, $\pi^+$, $K^+$ and $p$ --- which were produced and
identified in Pb-Pb collisions at 2.76 TeV in various
centrality intervals --- to the thermodynamically consistent Tsallis
distribution. The freezeout parameters in terms of the effective
temperature $T_{\rm eff}$, the non-extensitivity parameter $q$, and
the kinetic freezeout volume $V$ are extracted. These parameters
are obtained directly from the $p_T$ spectra.

In addition, the mean kinetic freezeout temperature $T_0$, the
mean transverse flow velocity $\beta_T$, the thermal temperature
$T_{\rm th}$ and the parameter $\zeta$ are extracted by an alternative
method. The behaviors of the above parameters are investigated
and it is observed that they all decrease
from central to peripheral collisions, except for
$q$, which has an opposite behavior.
The dependence of these parameters is also investigated
per unit of pseudorapidity,
\(\langle dN_{\rm ch}/d\eta\rangle\). We observed that
all these parameters increase with increasing 
\(\langle dN_{\rm ch}/d\eta\rangle\), again except for $q$.

$T_{\rm eff}$, $q$ and $V$ are mass dependent: the former is larger
for heavier particles, while the latter two are larger for lighter
particles. The mass dependence of effective temperature and kinetic
freezeout volume support the multiple kinetic
freezeout scenario, and the volume differential freezeout scenario,
respectively.

Furthermore, we observed a difference in the trend of $q$
for different particles with respect to centrality and
\(\langle dN_{\rm ch}/d\eta\rangle\). For instance, for $\pi^+$ it
increases continuously, while for $K^+$ it remains approximately constant
beyond a certain level of centrality. Similarly, for $p$ it increases up
to a certain centrality classes, then it decreases. Another similar behavior
is observed for $q$ with respect to \(\langle dN_{\rm ch}/d\eta\rangle\).

The higher temperature in central collisions  is because
central collisions experience a harsh squeeze, where the
degree of excitation of the system is high and so is the
temperature. In central systems, \(\langle dN_{\rm ch}/d\eta\rangle\)
is larger, which corresponds
to a hotter system because of the deposition of a huge amount
of energy that results in higher temperature at
the stage of kinetic freezeout, and ultimately in earlier
thermalization. Likewise, in central collisions, or at larger
\(\langle dN_{\rm ch}/d\eta\rangle\), the interactions among the particles
are frequent, which results in smaller values of $q$ that indicate that
the system is closer to equilibrium.

At last, $V$ decreases towards peripheral collisions,
since less particles interact, which leads to less binary
collisions in parton re-scattering. Thus the system has a steady
approach towards equilibrium, and the case of lower
\(\langle dN_{\rm ch}/d\eta\rangle\) is analogous.

When the \(\langle T_{\rm eff}\rangle\) and
\(\langle q \rangle\) dependencies on \(\langle N_{\rm part}\rangle\)
and \(\langle dN_{\rm ch}/d\eta\rangle\) are systematically compared
across various collision systems and energies, we observe fundamental
differences in the particle production mechanisms.
\(\langle T_{\rm eff}\rangle\) and \(\langle dN_{\rm ch}/d\eta\rangle\)
exhibit a high, positive correlation 
in heavy-ion collisions (Pb-Pb, Xe-Xe, and Au-Au), demonstrating the
critical role of the system size and energy density in attaining
thermalization.
This pattern is most prominent at LHC energy, where the largest
\(\langle T_{\rm eff}\rangle\) values are found in central Pb-Pb
collisions.
Small systems, such as p-Pb and p-p, on the other hand, reveal weaker
dependencies, with \(\langle T_{\rm eff}\rangle\) staying nearly
constant over the measured ranges of \(\langle N_{\rm part}\rangle\)
and \(\langle dN_{\rm ch}/d\eta\rangle\).

The non-extensitivity parameter \(\la q\ra \), however, exhibits
the opposite trends, decreasing as \(\langle N_{\rm part}\rangle\) and
\(\langle dN_{\rm ch}/d\eta\rangle\) increase in heavy-ion systems.
This indicates that thermal equilibrium in central collisions will
be achieved gradually. For the highest multiplicities and participant
numbers, this trend is particularly noticeable in Pb-Pb collisions
at LHC energy, when \(\langle q\rangle\) approaches unity. However,
small systems continue to have high \(\langle q\rangle\)-values in
all measured \(\langle N_{\rm part}\rangle\) and
\(\langle dN_{\rm ch}/d\eta\rangle\) ranges, which is indicative
of their non-extensive nature. Strong evidence for the shift
from non-thermal dynamics in small systems to a thermalized
behavior in large, high-multiplicity collisions is provided by
the anti-correlation between \(\langle T_{\rm eff}\rangle\)
and \(\langle q\rangle\), which is generically observed
across collision systems and energies.
The degree of thermalization is determined by the system size (via
\(\langle N_{\rm part}\rangle\)) and by the energy density (reflected
in \(\langle dN_{\rm ch}/d\eta\rangle\)).

We also observed that the parameter $\zeta$ decreases from central
to peripheral collisions because of reduced interactions and
fluctuations in the generated medium towards the periphery.
\\

\noindent
{\bf Acknowledgments:}
The authors extend their appreciation to the Deanship of Scientific
Research at Northern Border University, Arar, KSA for funding this
research work through the project number NBU-FFR-2024-2461-05.
This work was also funded by Princess Nourah bint Abdulrahman
University Researchers Supporting Project number (PNURSP2025R106),
Princess Nourah bint Abdulrahman University, Riyadh, Saudi Arabia,
and Ajman University Internal Research Grant No.\
[DRGS Ref. 2025-IRG-HBS-13], and by the Mexican funding agency
UNAM-DGAPA through project PAPIIT IG100322.
\\
\end{multicols}

{\small
\begin{thebibliography}{99}
\setlength{\itemsep}{-1pt}
\bibitem{QGPLQCD}
S.~D.\ Katz, S.\ Krieg, C.\ Ratti, K.~K.\ Szabo,
Is there still any $T_{c}$ mystery in lattice QCD?
Results with physical masses in the continuum limit III,
\href{https://doi.org/10.1007/JHEP09(2010)073}
{{\it JHEP} 1009 (2010) 073}
arXiv:1005.3508 [hep-lat]. \\
T.\ Bhattacharya {\it et al.}\ (HotQCD Collaboration),
QCD Phase Transition with Chiral Quarks and Physical Quark Masses,
\href{https://doi.org/10.1103/PhysRevLett.113.082001}
{{\it Phys.\ Rev.\ Lett.}\ {\bf 113} (2014) 082001}
arXiv:1402.5175 [hep-lat].

\bibitem{Srivastava:2016ayf}
E.\ Shuryak, "Quark-Gluon Plasma, Heavy Ion Collisions and Hadrons",
World Scientific Lecture Notes in Physics, 85 (2024).

\bibitem{Retiere:2003kf} F.~Reti\`ere and M.~A.~Lisa,
\href{https://doi.org/10.1103/PhysRevC.70.044907}
{Phys.\ Rev.\ C \textbf{70}, 044907 (2004)}
[arXiv:nucl-th/0312024 [nucl-th]].

\bibitem{Gale:2013da} C.~Gale, S.~Jeon and B.~Schenke,
\href{https://doi.org/10.1142/S0217751X13400113}
{Int.\ J.\ Mod.\ Phys.\ A \textbf{28}, 1340011 (2013)}
[arXiv:1301.5893 [nucl-th]].

\bibitem{Heinz:2013th} U.~Heinz and R.~Snellings,
\href{https://doi.org/10.1146/annurev-nucl-102212-170540}
{Ann.\ Rev.\ Nucl.\ Part.\ Sci.\ \textbf{63}, 123-151 (2013)}
[arXiv:1301.2826 [nucl-th]].

\bibitem{Bjorken:1982qr} J.~D.~Bjorken,
\href{https://doi.org/10.1103/PhysRevD.27.140}
{Phys.\ Rev.\ D \textbf{27}, 140-151 (1983)}.

\bibitem{aaa}
R.\ Vogt, in: Ultrarelativistic Heavy-Ion Collisions, edited
by R.\ Vogt (Elsevier Science B.V., Amsterdam, 2007) pp. 221-278.
 
\bibitem{Becattini:1997uf} F.~Becattini,
\href{https://doi.org/10.1088/0954-3899/23/12/017}
{J.\ Phys.\ G \textbf{23}, 1933-1940 (1997)}
[arXiv:hep-ph/9708248 [hep-ph]].

\bibitem{Becattini:1995if} F.~Becattini,
\href{https://doi.org/10.1007/BF02907431}
{Z.\ Phys.\ C \textbf{69}, no.3, 485-492 (1996)}.

\bibitem{Feal:2020myr}
X.~Feal, C.~Pajares and R.~Vazquez,
\href{https://doi.org/10.1103/PhysRevC.104.044904}
{Phys.\ Rev.\ C \textbf{104}, no.4, 044904 (2021)}
[arXiv:2012.02894 [hep-ph]].

\bibitem{Braun-Munzinger:1994ewq}
P.~Braun-Munzinger, J.~Stachel, J.~P.~Wessels and N.~Xu,
\href{https://doi.org/10.1016/0370-2693(94)01534-J}
{Phys.\ Lett.\ B \textbf{344}, 43-48 (1995)}
[arXiv:nucl-th/9410026 [nucl-th]].

\bibitem{Braun-Munzinger:2003htr}
P.~Braun-Munzinger, J.~Stachel and C.~Wetterich,
\href{https://doi.org/10.1016/j.physletb.2004.05.081}
{Phys.\ Lett.\ B \textbf{596}, 61-69 (2004)}
[arXiv:nucl-th/0311005 [nucl-th]].

\bibitem{Bialas:2015pla} A.~Bialas,
\href{https://doi.org/10.1016/j.physletb.2015.05.076}
{Phys.\ Lett.\ B \textbf{747}, 190-192 (2015)}
[arXiv:1506.00239 [hep-ph]].

\bibitem{Hagedorn:1983wk} R.~Hagedorn,
\href{https://doi.org/10.1007/BF02740917}
{Riv.\ Nuovo Cim.\ \textbf{6N10}, 1-50 (1983)}.

\bibitem{Hagedorn:1967dia} R.~Hagedorn,
\href{https://link.springer.com/article/10.1007/BF02755235}
{Nuovo Cim.\ A \textbf{52}, no.4, 1336-1340 (1967)}.

\bibitem{Hagedorn:1964zz} R.~Hagedorn,
in:
\href{https://doi.org/10.1007/978-3-319-17545-4_19}
{Rafelski, J. (eds) Melting Hadrons, Boiling Quarks --
From Hagedorn Temperature to Ultra-Relativistic Heavy-Ion
Collisions at CERN, Springer}.

\bibitem{Wilk:1999dr} G.~Wilk and Z.~Wlodarczyk,
\href{https://doi.org/10.1103/PhysRevLett.84.2770}
{Phys.\ Rev.\ Lett.\ \textbf{84}, 2770 (2000)}
[arXiv:hep-ph/9908459 [hep-ph]].

\bibitem{Biro:2020kve} G.~B\'\i{}r\'o, G.~G.~Barnaf\"oldi and T.~S.~Bir\'o,
\href{https://doi.org/10.1088/1361-6471/ab8dcb}
{J.\ Phys.\ G \textbf{47}, no.10, 105002 (2020)}
[arXiv:2003.03278 [hep-ph]].

\bibitem{Saraswat:2017kpg} K.~Saraswat, P.~Shukla and V.~Singh,
\href{https://doi.org/10.1088/2399-6528/aab00f}
{J.\ Phys.\ Comm.\ \textbf{2}, no.3, 035003 (2018)}
[arXiv:1706.04860 [hep-ph]].

\bibitem{Broniowski:2001we} W.~Broniowski and W.~Florkowski,
\href{https://doi.org/10.1103/PhysRevLett.87.272302}
{Phys.\ Rev.\ Lett.\ \textbf{87}, 272302 (2001)}
[arXiv:nucl-th/0106050 [nucl-th]].

\bibitem{Che:2020fbz} G.~Che, J.~Gu, W.~Zhang and H.~Zheng,
\href{https://doi.org/10.1088/1361-6471/ac09dc}
{J.\ Phys.\ G \textbf{48}, no.9, 095103 (2021)}
[arXiv:2010.14880 [nucl-th]].

\bibitem{Waqas:2024qlo} M.~Waqas, B.~Hassan, A.~Alnakhlani, M.~Ajaz,
  A.~Altalbe, R.~Ghodhbani and A.~Haj Ismail,
\href{https://doi.org/10.1016/j.rinp.2024.107894}
{Results Phys.\ \textbf{64}, 107894 (2024)}.

\bibitem{Waqas:2019mjp} M.~Waqas and B.~C.~Li,
\href{https://doi.org/10.1155/2020/1787183}
{Adv.\ High Energy Phys. \textbf{2020}, 1787183 (2020)}
[arXiv:1909.11339 [hep-ph]].

\bibitem{STAR:2017sal} L.~Adamczyk \textit{et al.} [STAR],
\href{https://doi.org/10.1103/PhysRevC.96.044904}
{Phys.\ Rev.\ C \textbf{96}, no.4, 044904 (2017)}
[arXiv:1701.07065 [nucl-ex]].

\bibitem{Waqas:2021qkr} M.~Waqas and G.~X.~Peng,
\href{https://doi.org/10.1155/2021/6674470}
{Adv.\ High Energy Phys. \textbf{2021}, 6674470 (2021)}
[arXiv:2103.07852 [hep-ph]].

\bibitem{Wang:2023rpd}
R.~Q.~ Wang, Y.~H.~Li, J.~Song and F.~L.~Shao,
\href{https://doi.org/10.1103/PhysRevC.109.034907}
{Phys.\ Rev.\ C \textbf{109}, no.3, 034907 (2024)}
[arXiv:2309.16296 [nucl-th]].

\bibitem{Sharma:2024nkt} R.~Sharma, K.~Gopal, S.~R.~Sharma and C.~Jena,
\href{https://doi.org/10.48550/arXiv.2401.13629}
{arXiv:2401.13629 [hep-ph]}.

\bibitem{ALICE:2013mez} B.~Abelev \textit{et al.} [ALICE],
\href{https://doi.org/10.1103/PhysRevC.88.044910}
{Phys.\ Rev.\ C \textbf{88}, 044910 (2013)}
[arXiv:1303.0737 [hep-ex]].

\bibitem{CMS:2016zzh} V.~Khachatryan \textit{et al.} [CMS],
\href{https://doi.org/10.1016/j.physletb.2013.11.020}
{Phys.\ Lett.\ B \textbf{768}, 103-129 (2017)}
[arXiv:1605.06699 [nucl-ex]].

\bibitem{ALICE:2013wgn} B.~B.~Abelev \textit{et al.} [ALICE],
\href{https://doi.org/10.1016/j.physletb.2013.11.020}
{Phys.\ Lett.\ B \textbf{728}, 25-38 (2014)}
[arXiv:1307.6796 [nucl-ex]].

\bibitem{Schnedermann:1993ws} E.~Schnedermann, J.~Sollfrank and U.~W.~Heinz,
\href{https://doi.org/10.1103/PhysRevC.48.2462}
{Phys.\ Rev.\ C \textbf{48}, 2462-2475 (1993)}
[arXiv:nucl-th/9307020 [nucl-th]].

\bibitem{STAR:2006nmo} B.~I.~Abelev \textit{et al.} [STAR],
\href{https://doi.org/10.1103/PhysRevC.75.064901}
{Phys.\ Rev.\ C \textbf{75}, 064901 (2007)}
[arXiv:nucl-ex/0607033 [nucl-ex]].

\bibitem{UA1:1982fux} G.~Arnison \textit{et al.} [UA1],
\href{https://doi.org/10.1016/0370-2693(82)90623-2}
{Phys.\ Lett.\ B \textbf{118}, 167-172 (1982)}.

\bibitem{Cleymans:2011in} J.~Cleymans and D.~Worku,
\href{https://doi.org/10.1088/0954-3899/39/2/025006}
{J.\ Phys.\ G \textbf{39}, 025006 (2012)}
[arXiv:1110.5526 [hep-ph]].

\bibitem{Pereira:2007hp} F.~I.~M.~Pereira, R.~Silva and J.~S.~Alcaniz,
\href{https://doi.org/10.1103/PhysRevC.76.015201}
{Phys.\ Rev.\ C \textbf{76}, 015201 (2007)}
[arXiv:0705.0300 [nucl-th]].

\bibitem{Conroy:2010wt} J.~M.~Conroy, H.~G.~Miller and A.~R.~Plastino,
\href{https://doi.org/10.1016/j.physleta.2010.09.038}
{Phys.\ Lett.\ A \textbf{374}, 4581-4584 (2010)}
[arXiv:1006.3963 [cond-mat.stat-mech]].

\bibitem{ALICE:2013txf} B.~B.~Abelev \textit{et al.} [ALICE],
\href{https://doi.org/10.1140/epjc/s10052-013-2662-9}
{Eur.\ Phys.\ J.\ C \textbf{73}, no.12, 2662 (2013)}
[arXiv:1307.1093 [nucl-ex]].

\bibitem{CMS:2010tjh} V.~Khachatryan \textit{et al.} [CMS],
\href{https://doi.org/10.1103/PhysRevLett.105.022002}
{Phys.\ Rev.\ Lett.\ \textbf{105}, 022002 (2010)}
[arXiv:1005.3299 [hep-ex]].

\bibitem{ATLAS:2010jvh} G.~Aad \textit{et al.} [ATLAS],
\href{https://doi.org/10.1088/1367-2630/13/5/053033}
{New J.\ Phys.\ \textbf{13}, 053033 (2011)}
[arXiv:1012.5104 [hep-ex]].

\bibitem{PHENIX:2011rvu} A.~Adare \textit{et al.} [PHENIX],
\href{https://doi.org/10.1103/PhysRevC.83.064903}
{Phys.\ Rev.\ C \textbf{83}, 064903 (2011)}
[arXiv:1102.0753 [nucl-ex]].

\bibitem{ALICE:2010syw} K.~Aamodt \textit{et al.} [ALICE],
\href{https://doi.org/10.1016/j.physletb.2010.08.026}
{Phys.\ Lett.\ B \textbf{693}, 53-68 (2010)}
[arXiv:1007.0719 [hep-ex]].

\bibitem{Biro:2012fiy} T.~S.~Bir\'o, G.~G.~Barnaf\"oldi and P.~Van,
\href{https://doi.org/10.1140/epja/i2013-13110-0}
{Eur.\ Phys.\ J. A \textbf{49}, 110 (2013)}
[arXiv:1208.2533 [hep-ph]].

\bibitem{Wilk:2012zn} G.~Wilk and Z.~Wlodarczyk,
\href{https://doi.org/10.1140/epja/i2012-12161-y}
{Eur.\ Phys.\ J.\ A \textbf{48}, 161 (2012)}
[arXiv:1203.4452 [hep-ph]].

\bibitem{PHENIX:2008psu} A.~Adare \textit{et al.} [PHENIX],
\href{https://doi.org/10.1103/PhysRevC.78.044902}
{Phys.\ Rev.\ C \textbf{78}, 044902 (2008)}
[arXiv:0805.1521 [nucl-ex]].

\bibitem{Biro:2014fna} 
T.~S.~Bir\'o, P.~Van, G.~G.~Barnaf\"oldi and K.~\"Urm\"ossy,
\href{https://doi.org/10.3390/e16126497}
{Entropy \textbf{16} 6497-6514 (2014)}
[arXiv:1409.5975 [cond-mat.stat-mech]].

\bibitem{Grosse-Oetringhaus:2019axb} J.~F.~Grosse-Oetringhaus,
\href{https://doi.org/10.22323/1.364.0711}
{PoS \textbf{EPS-HEP2019}, 711 (2020)}
[arXiv:2001.02880 [nucl-ex]].

\bibitem{ALICE:2016fzo} J.~Adam \textit{et al.} [ALICE],
\href{https://doi.org/10.1038/nphys4111}
{Nature Phys. \textbf{13}, 535-539 (2017)}
[arXiv:1606.07424 [nucl-ex]].

\bibitem{ATLAS:2012cix} G.~Aad \textit{et al.} [ATLAS],
\href{https://doi.org/10.1103/PhysRevLett.110.182302}
{Phys.\ Rev.\ Lett.\ \textbf{110}, no.18, 182302 (2013)}
[arXiv:1212.5198 [hep-ex]].

\bibitem{Mishra:2018pio} A.~N.~Mishra, A.~Ortiz and G.~Pai\'{c},
\href{https://doi.org/10.1103/PhysRevC.99.034911}
{Phys.\ Rev.\ C \textbf{99}, no.3, 034911 (2019)}
[arXiv:1805.04572 [hep-ph]].

\bibitem{Nath:2019mkf} A.~N.~Mishra and G.~Pai\'{c},
\href{https://doi.org/10.48550/arXiv.1905.06918}
{arXiv:1905.06918 [hep-ph]}.

\bibitem{Zaccolo:2015udc} V.~Zaccolo [ALICE],
\href{https://doi.org/10.1016/j.nuclphysa.2016.01.025}
{Nucl.\ Phys.\ A \textbf{956}, 529-532 (2016)}
[arXiv:1512.05273 [hep-ex]].

\bibitem{Csorgo:1994fg} T.~Cs\"org\"o, B.~L\"orstad and J.~Zim\'anyi,
\href{https://doi.org/10.1016/0370-2693(94)91356-0}
{Phys.\ Lett.\ B \textbf{338}, 134-140 (1994)}
[arXiv:nucl-th/9408022 [nucl-th]].

\bibitem{Helgesson:1997zz}
J.~Helgesson, T.~Cs\"org\"o, M.~Asakawa and B.~L\"orstad,
\href{https://doi.org/10.1103/PhysRevC.56.2626}
{Phys.\ Rev.\ C \textbf{56}, 2626-2635 (1997)}
[arXiv:nucl-th/9506006].

\bibitem{Waqas:2020ioh} M.~Waqas, F.~H.~Liu, L.~L.~Li and H.~M.~Alfanda,
\href{https://doi.org/10.1007/s41365-020-00821-7}
{Nucl.\ Sci.\ Tech.\ \textbf{31}, no.11, 109 (2020)}
[arXiv:2001.06796 [hep-ph]].

\bibitem{Moore:2004tg} G.~D.~Moore and D.~Teaney,
\href{https://doi.org/10.1103/PhysRevC.71.064904}
{Phys.\ Rev.\ C \textbf{71}, 064904 (2005)}
[arXiv:hep-ph/0412346 [hep-ph]].

\bibitem{Csorgo:2001xm}
T.~Cs\"org\"o, S.~V.~Akkelin, Y.~Hama, B.~Luk\'acs and Y.~M.~Sinyukov,
\href{https://doi.org/10.1103/PhysRevC.67.034904}
{Phys.\ Rev.\ C \textbf{67}, 034904 (2003)}
[arXiv:hep-ph/0108067 [hep-ph]].

\bibitem{PHENIX:2011rvu} A.~Adare \textit{et al.} [PHENIX],
\href{https://doi.org/10.1103/PhysRevC.83.064903}
{Phys.\ Rev.\ C \textbf{83}, 064903 (2011)}
[arXiv:1102.0753 [nucl-ex]].

\bibitem{Wei:2016ihj} H.~R.~Wei, F.~H.~Liu and R.~A.~Lacey,
\href{https://doi.org/10.1140/epja/i2016-16102-6}
{Eur.\ Phys.\ J.\ A \textbf{52}, no.4, 102 (2016)}
[arXiv:1601.07045 [hep-ph]].

\bibitem{Tang:2008ud}
Z.~Tang, Y.~Xu, L.~Ruan, G.~van Buren, F.~Wang and Z.~Xu,
\href{doi:10.1103/PhysRevC.79.051901}
{Phys. Rev. C \textbf{79}, 051901 (2009)}
[arXiv:0812.1609 [nucl-ex]].

\bibitem{Chatterjee:2014lfa} S.~Chatterjee, B.~Mohanty and R.~Singh,
\href{https://doi:10.1103/PhysRevC.92.024917}
{Phys. Rev. C \textbf{92}, no.2, 024917 (2015)}
[arXiv:1411.1718 [nucl-th]].

\bibitem{Chatterjee:2015fua} S.~Chatterjee, S.~Das, L.~Kumar,
D.~Mishra, B.~Mohanty, R.~Sahoo and N.~Sharma,
\href{https://doi:10.1155/2015/349013}
{Adv. High Energy Phys. \textbf{2015}, 349013 (2015)}.

\bibitem{Thakur:2016boy}
D.~Thakur, S.~Tripathy, P.~Garg, R.~Sahoo and J.~Cleymans,
\href{https://doi:10.1155/2016/4149352}
{Adv. High Energy Phys. \textbf{2016}, 4149352 (2016)}
[arXiv:1601.05223 [hep-ph]].

\bibitem{Waqas:2021rmb} M.~Waqas, G.~X.~Peng and F.~H.~Liu,
\href{https://doi:10.1088/1361-6471/abdd8d}
{J. Phys. G \textbf{48}, no.7, 075108 (2021)}
[arXiv:2101.07971 [hep-ph]].

\bibitem{ALICE:2019hno} S.~Acharya \textit{et al.} [ALICE],
\href{https://doi:10.1103/PhysRevC.101.044907}
{Phys. Rev. C \textbf{101}, no.4, 044907 (2020)}
[arXiv:1910.07678 [nucl-ex]].

\bibitem{Lao:2021wub} H.~L.~Lao, F.~H.~Liu and B.~Q.~Ma,
\href{https://doi:10.3390/e23070803}
{Entropy \textbf{23}, no.7, 803 (2021)}.

\bibitem{Lao:2017dtr}
H.~L.~Lao, F.~H.~Liu, B.~C.~Li, M.~Y.~Duan and R.~A.~Lacey,
\href{https://doi:10.1007/s41365-018-0504-z}
{Nucl. Sci. Tech. \textbf{29}, no.11, 164 (2018)}
[arXiv:1708.07749 [nucl-th]].

\bibitem{Waqas:2021nku}
M.~Waqas, G.~X.~Peng, R.~Q.~Wang, M.~Ajaz and A.~A.~Ismail,
\href{https://doi:10.1140/epjp/s13360-021-02089-1}
{Eur. Phys. J. Plus \textbf{136}, no.10, 1082 (2021)}
[arXiv:2110.09505 [nucl-th]].

\bibitem{Waqas:2018tkk}
M.~Waqas and F.~H.~Liu,
\href{https://doi:10.1007/s12648-021-02058-5}
{Indian J. Phys. \textbf{96}, no.4, 1217-1235 (2022)}
[arXiv:1806.05863 [hep-ph]].

\bibitem{Waqas:2022nae}
M.~Waqas, G.~X.~Peng, M.~Ajaz, A.~Haj Ismail and E.~A.~Dawi,
\href{https://doi.org/10.1103/PhysRevD.106.075009}
{Phys.\ Rev.\ D \textbf{106}, no.7, 075009 (2022)}
[arXiv:2209.07073 [hep-ph]].

\bibitem{Waqas:2018xrz}
M.~Waqas, F.~H.~Liu, S.~Fakhraddin and M.~A.~Rahim,
\href{https://doi.org/10.1007/s12648-019-01396-9}
{Indian J.\ Phys.\ \textbf{93}, no.10, 1329-1343 (2019)}
[arXiv:1806.04312 [nucl-th]].

\bibitem{Waqas:2020ygr} M.~Waqas, F.~H.~Liu, R.~Q.~Wang and I.~Siddique,
\href{https://doi.org/10.1140/epja/s10050-020-00192-y}
{Eur.\ Phys.\ J.\ A \textbf{56}, no.7, 188 (2020)}
[arXiv:2007.00825 [hep-ph]].

\bibitem{imran2025} A.~Rehman \textit{et al.},
\href{https://doi.org/10.1142/S0217732325500634}
{Mod.\ Phys.\ Lett.\ A \textbf{40}, no.19n20, 2550063 (2025)}.

\bibitem{ALICE:2015juo} J.~Adam \textit{et al.} [ALICE],
\href{https://doi:10.1103/PhysRevLett.116.222302}
{Phys. Rev. Lett. \textbf{116}, no.22, 222302 (2016)}
[arXiv:1512.06104 [nucl-ex]].

\bibitem{STAR:2008med} B.~I.~Abelev \textit{et al.} [STAR],
\href{https://doi:10.1103/PhysRevC.79.034909}
{Phys. Rev. C \textbf{79}, 034909 (2009)}
[arXiv:0808.2041 [nucl-ex]].

\bibitem{ALICE:2018pal} S.~Acharya \textit{et al.} [ALICE],
\href{https://doi:10.1103/PhysRevC.99.024906}
{Phys. Rev. C \textbf{99}, no.2, 024906 (2019)}
[arXiv:1807.11321 [nucl-ex]].

\bibitem{ALICE:2020nkc} S.~Acharya \textit{et al.} [ALICE],
\href{https://doi:10.1140/epjc/s10052-020-8125-1}
{Eur. Phys. J. C \textbf{80}, no.8, 693 (2020)}
[arXiv:2003.02394 [nucl-ex]].

\bibitem{ALICE:2018hza} S.~Acharya \textit{et al.} [ALICE],
\href{https://doi:10.1016/j.physletb.2018.10.052}
{Phys. Lett. B \textbf{788}, 166-179 (2019)}
[arXiv:1805.04399 [nucl-ex]].

\bibitem{BRAHMS:2016klg} I.~C.~Arsene \textit{et al.} [BRAHMS],
\href{https://doi:10.1103/PhysRevC.94.014907}
{Phys. Rev. C \textbf{94}, no.1, 014907 (2016)}
[arXiv:1602.01183 [nucl-ex]].

\bibitem{STAR:2007poe} B.~I.~Abelev \textit{et al.} [STAR],
\href{https://doi.org/10.48550/arXiv.nucl-ex/0703016}
{arXiv:nucl-ex/0703016 [nucl-ex]}.
  
\bibliographystyle{plain}
\end{thebibliography}}
\end{document}